\documentclass[aps,twocolumn,pra,eqsecnum, floats, showpacs, notitlepage, superscriptaddress,10pt]{revtex4-1}
\usepackage{graphicx}
\usepackage{dcolumn}
\usepackage{amsmath}
\usepackage{verbatim}
\usepackage{bbold}
\usepackage{subeqnarray}
\usepackage{bm}
\usepackage{color}
\usepackage{lmodern}
\usepackage[caption=false]{subfig}
\usepackage{amssymb}
\usepackage{mathrsfs}
\usepackage{pict2e}
\usepackage{tikz}
\usepackage{mathtools}

\DeclarePairedDelimiterX\braket[2]{\langle}{\rangle}{#1 \delimsize\vert #2}
\usepackage[breaklinks=true,colorlinks=true,linkcolor=blue,urlcolor=blue,citecolor=magenta]{hyperref}

 \newcounter{multifig}
 
\begin{document}
\title{Spontaneous emission in an exponential model}

\author{A. D. Kammogne}
\affiliation{Mesoscopic and Multilayers Structures Laboratory, Faculty of Science, Department of Physics, University of Dschang, Cameroon}

\author{L. C. Fai}
\affiliation{Mesoscopic and Multilayers Structures Laboratory, Faculty of Science, Department of Physics, University of Dschang, Cameroon}

\date{\today}

\begin{abstract}

The phenomenon of spontaneous emission can lead to the creation of an imaginary coupling and a shift. To explore this, we utilized the renormalized first Nikitin model, revealing an exponential detuning variation with a phase and an imaginary coupling along with the shift. By employing the time-dependent Schr\"odinger equation, we investigated the behavior of our system. Our findings indicate that the imaginary coupling provides specific information, while the shift generates allowed and forbidden zones in the energy diagram of the real part of the energy. In the diagram of the imaginary part of the energy, time dictates order or chaos in the system and identifies the information transmission zone. Notably, the first Nikitin model exhibits similarities to the Rabi model in the short-time approximation. Our theoretical conclusions are consistent with numerical solutions.

\end{abstract}
\maketitle

\section*{Introduction}\label{Sec1}

The concept of spontaneous emission underpins many physical phenomena, such as non-Hermiticity \cite{Ke, Hadad, Jakobczyk, Becker, F. Chen, Zongping, Khanbekyan, WenjieZhou, Satoshi, YimingHuang}, dissipation \cite{Ao, Kudlis, GSAgarwal, DMeschede, Ginzburg, Pelton, Søndergaard, TBrandes, Vasilii, Alejandro} between quantum systems, the production of lasers \cite{Qi, YYamamoto, CHenry, MarianoA, NPBarnes, AVBogatskaya, HCao, FHNicoll, WengW, YWHuang}, and various other natural occurrences. Non-Hermiticity has become a focal topic of scientific discussion worldwide and has garnered significant attention in the field of quantum mechanics. In some cases, spontaneous emission and non-Hermiticity are interrelated. For instance,  \cite{Siegman} demonstrated that excessive spontaneous emission in a laser oscillator increases noise in non-Hermitian optical systems. Furthermore, \cite{Ferrier} confirmed spontaneous emission at exceptional points through photoluminescence measurements, while \cite{Pick} developed a general theory of spontaneous emission, highlighting its applicability to dispersive media.

Several authors suggest that the chirality of non-Hermitian models, such as the SSH (Su-Schrieffer-Heeger) model \cite{Lieu}, generates a complex Berry phase, explaining the presence of gapless edge modes. Yuto Ashida et al.  \cite{Ashida} explored the topology of non-Hermitian systems, introducing the concept of band topology in complex spectra and defining their classifications. Another study \cite{Lee} showed that spontaneous decay induces quantum phase transitions in non-Hermitian systems, revealing a connection between dissipation and non-Hermitian properties in quantum systems. Non-Hermitian systems are characterized by complex eigenenergies comprising real and imaginary components. This notion can be counterintuitive because quantum mechanics dictates that a system must be Hermitian, with real eigenenergies, to measure observables.

\begin{figure}[h!]
	\centering
	\includegraphics[width=0.32\textwidth, height =0.32\textwidth]{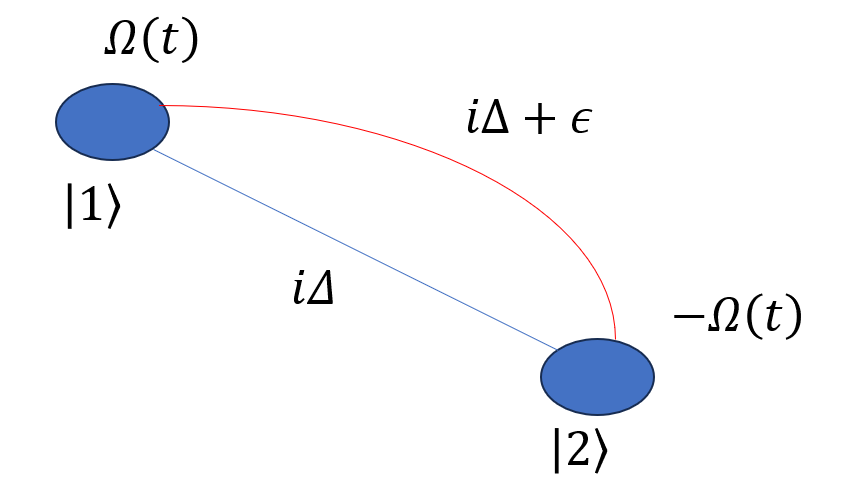}
	\caption{\small Spontaneous emission between two diabatic states during their interaction. In this process, the states $\left| 1 \right\rangle$ and $\left| 2 \right\rangle$ are associated with the detuning $\Omega \left( t \right)$ and $-\Omega \left( t \right)$  Injection of the magnetic field is at the origin of the creation of spontaneous emission characterized by an imaginary coupling followed by a shift. This reflects the non-hermitian character of our Hamiltonian}.  \label{fig1}
\end{figure}

In studying the interaction between a two-level system and its environment, we propose a new scenario where spontaneous emission generates imaginary coupling, as illustrated in Fig.\ref{fig1} accompanied by a static shift. In this context, we consider two quantum states, denoted as $\left| 1 \right\rangle $ and $\left| 2 \right\rangle $,  represented respectively by positive detuning $\Omega \left( t \right)$ and negative detuning $- \Omega \left( t \right)$. These states are coupled via an imaginary term $i\Delta$ and the static shift.

To investigate the dynamics of this scenario, we selected the exponential Nikitin model, characterized by exponentially varying detuning, imaginary coupling, and a static shift term. The imaginary term primarily drives the non-Hermitian nature of our Hamiltonian. While quantum mechanics generally permits the measurement of observables only when the spectrum is real, this work examines how complex spectra emerge when coupling becomes complex. The Nikitin model has been extensively explored in previous works, such as those by Nesbitt et al.  \cite{Nesbitt} and Vitanov \cite{Vitanov},  who highlighted its similarity to the Demkov-Kunike model under strong coupling. The Nikitin model also describes decay processes \cite{Kammogne1}  and has applications in interferometry \cite{Kammogne2}.  Furthermore, collision dynamics between two states have been described using an exponential model by V. Sidis et al. \cite{Sidis}.

In this work, we present a novel perspective on spontaneous emission, incorporating imaginary coupling with a shift in the off-diagonal Hamiltonian terms. Spontaneous emission plays a crucial role in Bose-Einstein condensates \cite{Li, Radka, MRHush, CMSavage, DCDai, DSnoke, LWClark, JKasprzak,  JulianSchmitt, HuiDeng, J.Bloch, Heinzen}, radio-active process \cite{Reisenbauer, HJRose, JGross, HJWALKE, FHartmann-Boutron, Brakel, Goldhaber, Gromov, Poenaru, EPickup} and ultra-cold trapped gas \cite{Shi, TimothyP, Julienne1, Ostermann, Julienne, Hilico, DylanJervis, IBloch, Roux, TEDrake, LSantos}. Our aim is to construct a theoretical framework to better understand this model's behavior and elucidate the role of imaginary coupling.

This paper is organized as follows: in the first section, we present the exponential Nikitin model. In the second section, we solve the time-dependent Schr\"odinger equation to analyze the model's dynamics. In the third section, we discuss the energy spectrum. Finally, we summarize our findings and propose future research directions.

\section{Exponential model}\label{sec2}

To describe the interaction between diabatic states $\left| 1 \right\rangle $ and $\left| 2 \right\rangle$ during spontaneous emission, we select the first exponential Nikitin model due to its exponentially varying parameters.

Our model is described by the Hamiltonian:
\begin{equation}
	H\left( t \right) = \Omega \left( t \right){\sigma _z} + \delta {\sigma _x},\label{1.1}
\end{equation}

where $\Omega \left (t \right)$ and $\delta $ represent, respectively, the detuning and the Rabi frequency. Each parameter plays a crucial role in describing the model. The detuning $\Omega \left( t \right)$ determines the direction of the magnetic field via the Pauli matrix $\,{\sigma _z}$, while the Rabi frequency measures the coupling strength in the transverse direction $\,{\sigma _x}$. These quantities are defined as follows:
\begin{equation}
	\Omega \left( t \right) = \frac{1}{2}\left( {A\exp \left( {\alpha t + \beta } \right) + \epsilon } \right), \label{1.2}
\end{equation}

\begin{equation}
	\delta = \frac{1}{2}\left( i\Delta  + \epsilon \right). \label{1.3}
\end{equation}

As shown, the detuning $\Omega \left( t \right)$ varies exponentially, as observed in \cite{Kammogne1},  but with the additional feature of a phase $\beta $. Here, $A$ represents the amplitude, responsible for the signal strength; $\alpha$ is the sweep velocity induced by an external field; and $\beta$ is the phase, which controls oscillations' increase or decrease. The coupling $\delta$ includes an imaginary component due to the effects of spontaneous emission, accompanied by a shift $\epsilon$. At this stage, the shift acts as the real coupling responsible for the interaction between the levels. The Pauli matrices ${\sigma _\omega }$ with $\omega  = x,y,z$ are defined as the generators of the $SU (2)$ Lie algebra describing the dynamics of the two-level system.

Compared to the work of \cite{Kammogne1}, which studied the Nikitin model with real detuning and coupling in the context of decay, this work focuses on spontaneous emission with imaginary coupling. The combination of imaginary coupling and the shift introduces nonlinearity to our system.

\section{Methods}\label{sec3}

Among the various methods available to study the dynamics of this model-such as the Bloch picture \cite{Zlatanov, Tchapda} with the Lindbard equation \cite{Cortes} , the Feymann Path integral \cite{Bagarello}, the Schr\"odinger picture with the time-dependent Schr\"odinger \cite{Kammogne1}, and so on. we chose the Schr\"odinger picture. This method provides a wealth of information, including probability amplitudes, the propagator, and the transition probabilities.

\subsection{Schr\"odinger picture}

The dynamics of our model are determined using the time-dependent Schr\"odinger equation, which establishes the relationship between probabilities and their evolution through the matrix elements of the propagator. The non-Hermitian nature of our system arises from the effects of spontaneous emission, introducing an imaginary coupling. The Schr\"odinger equation is expressed as:
\begin{equation}
	i\frac{d}{{dt}}C\left( t \right) = H\left( t \right)C\left( t \right).
\end{equation}

Using an appropriate gauge transformation ${C_1}\left( t \right) = {\psi _1}\left( t \right)\exp \left( -{\frac{i}{2}\int\limits_{{t_0}}^t {\Omega \left( {{t_1}} \right)d{t_1}} } \right)$, and a change of variables $x\left( t \right) = \exp \left( {\alpha t + \beta } \right)$, the differential equation transforms into:

\begin{figure}[h!]
	\centering
	\includegraphics[width=0.22\textwidth, height =0.22\textwidth]{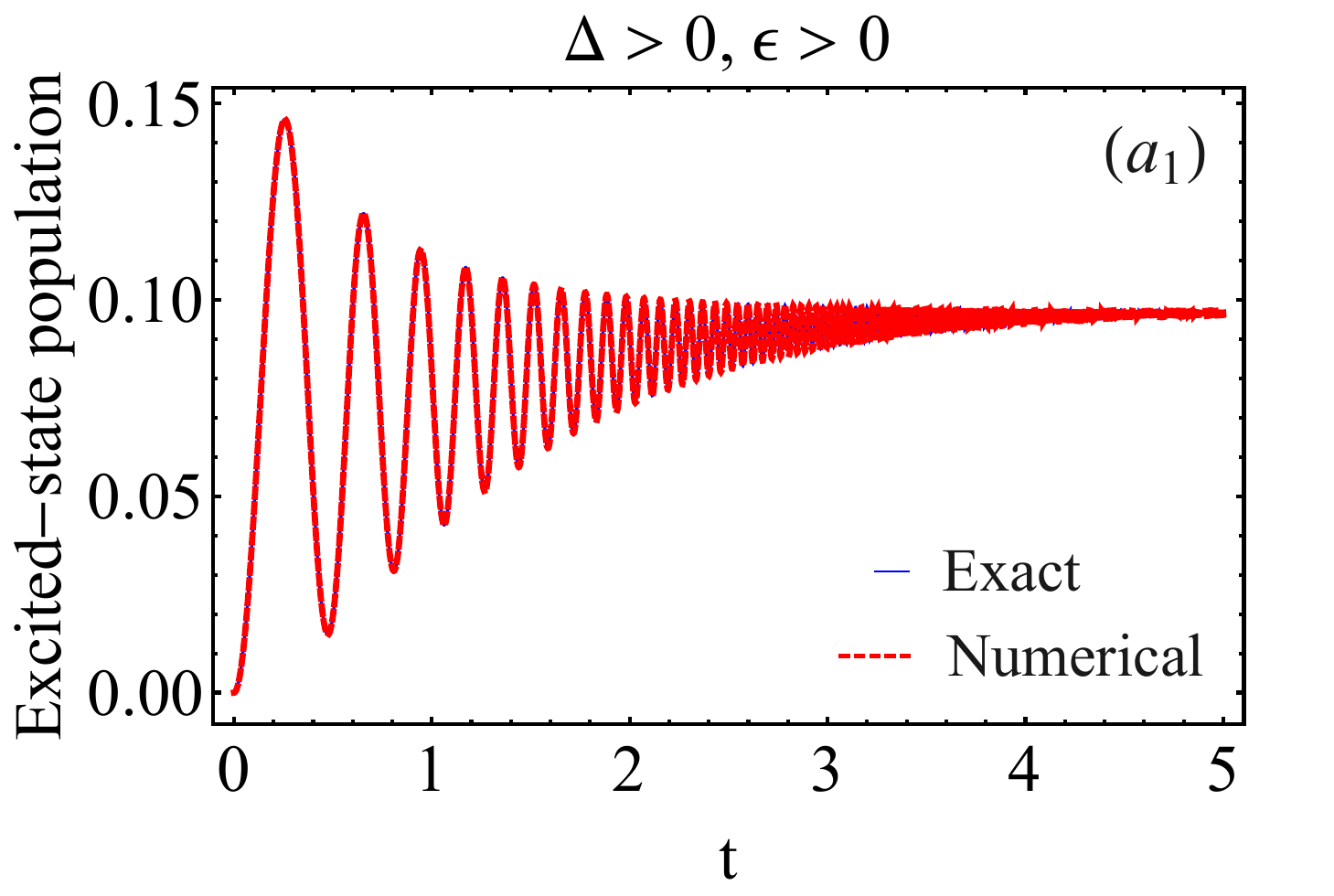}
	\includegraphics[width=0.22\textwidth, height =0.22\textwidth]{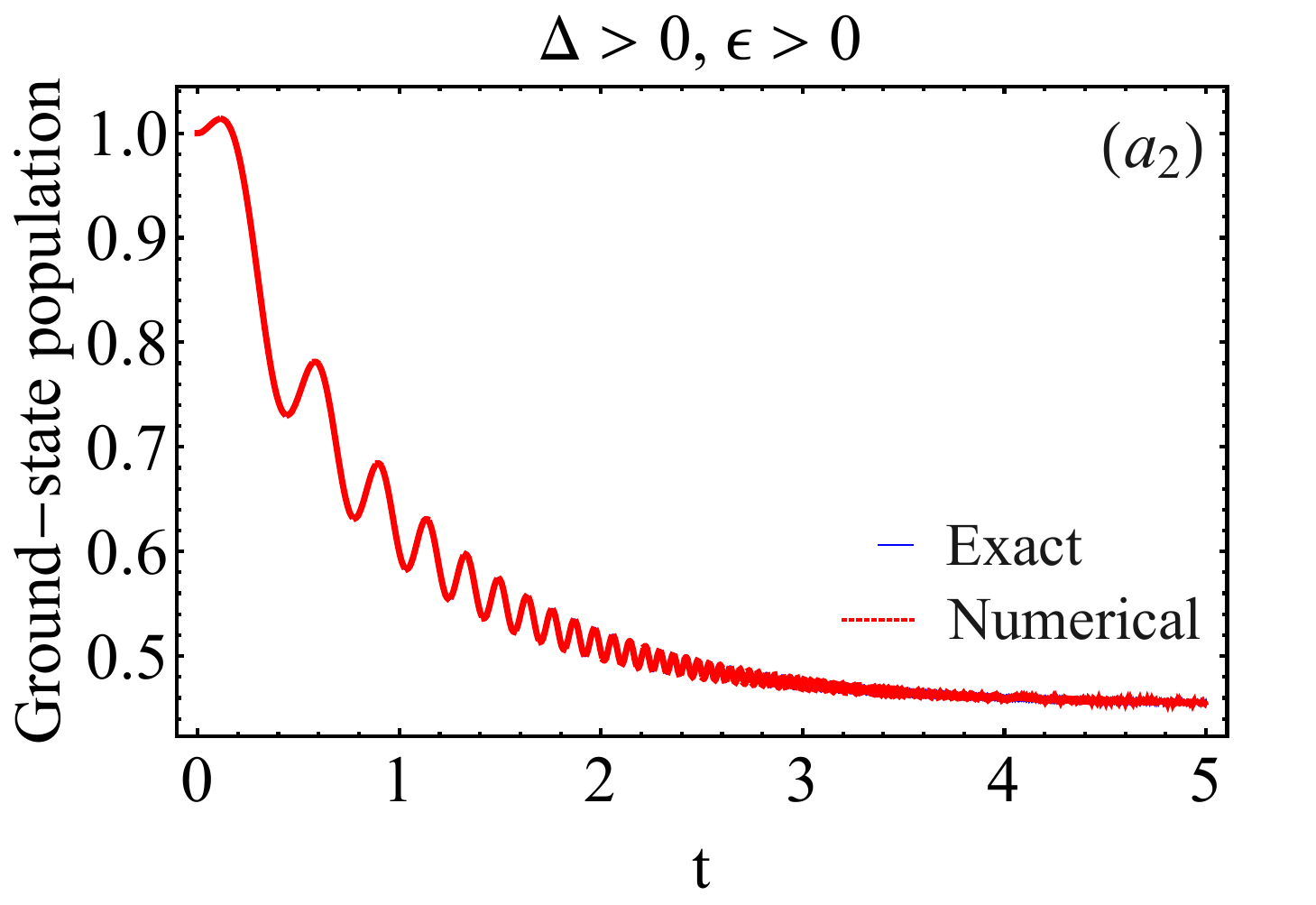}
	\includegraphics[width=0.22\textwidth, height =0.22\textwidth]{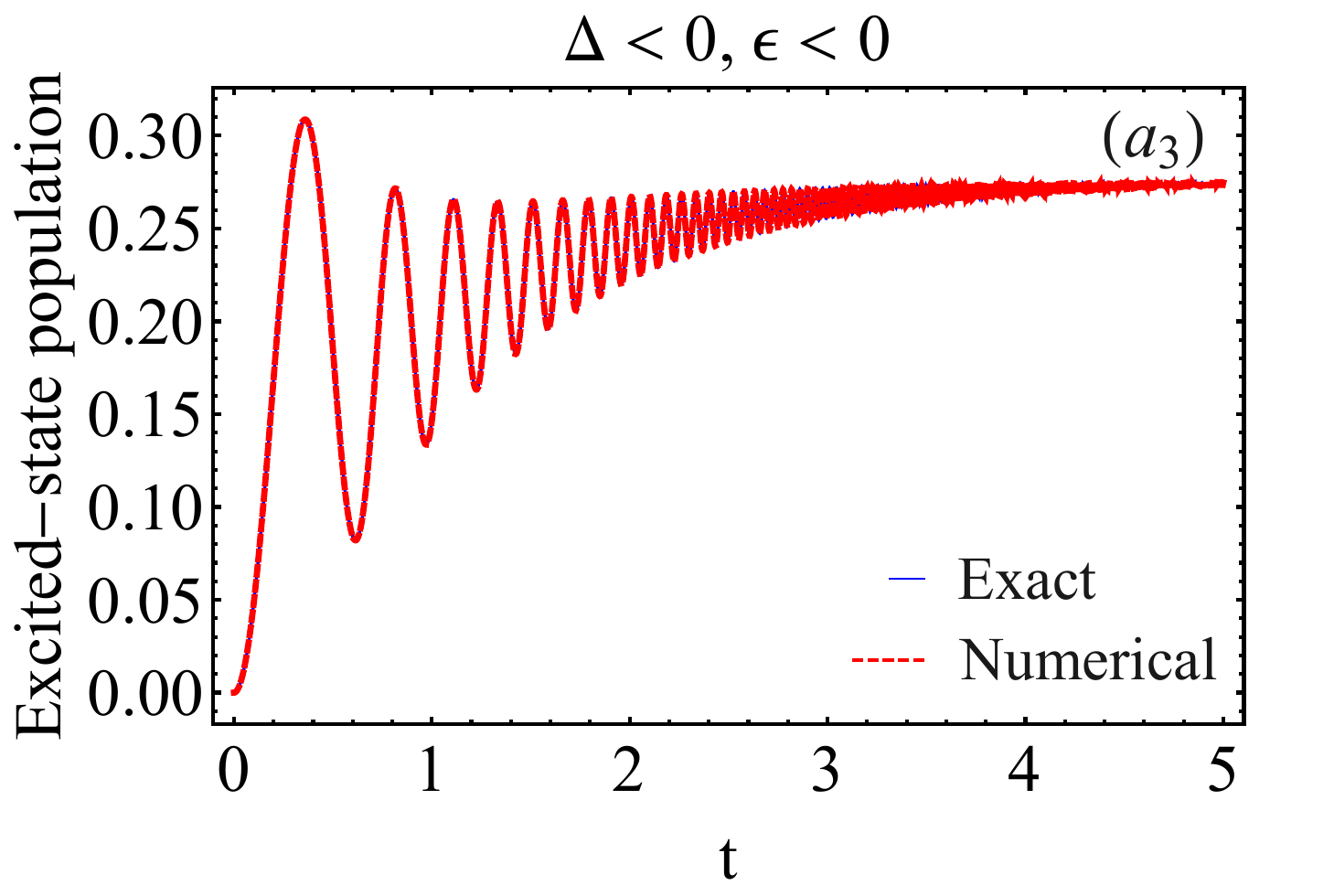}
	\includegraphics[width=0.22\textwidth, height =0.22\textwidth]{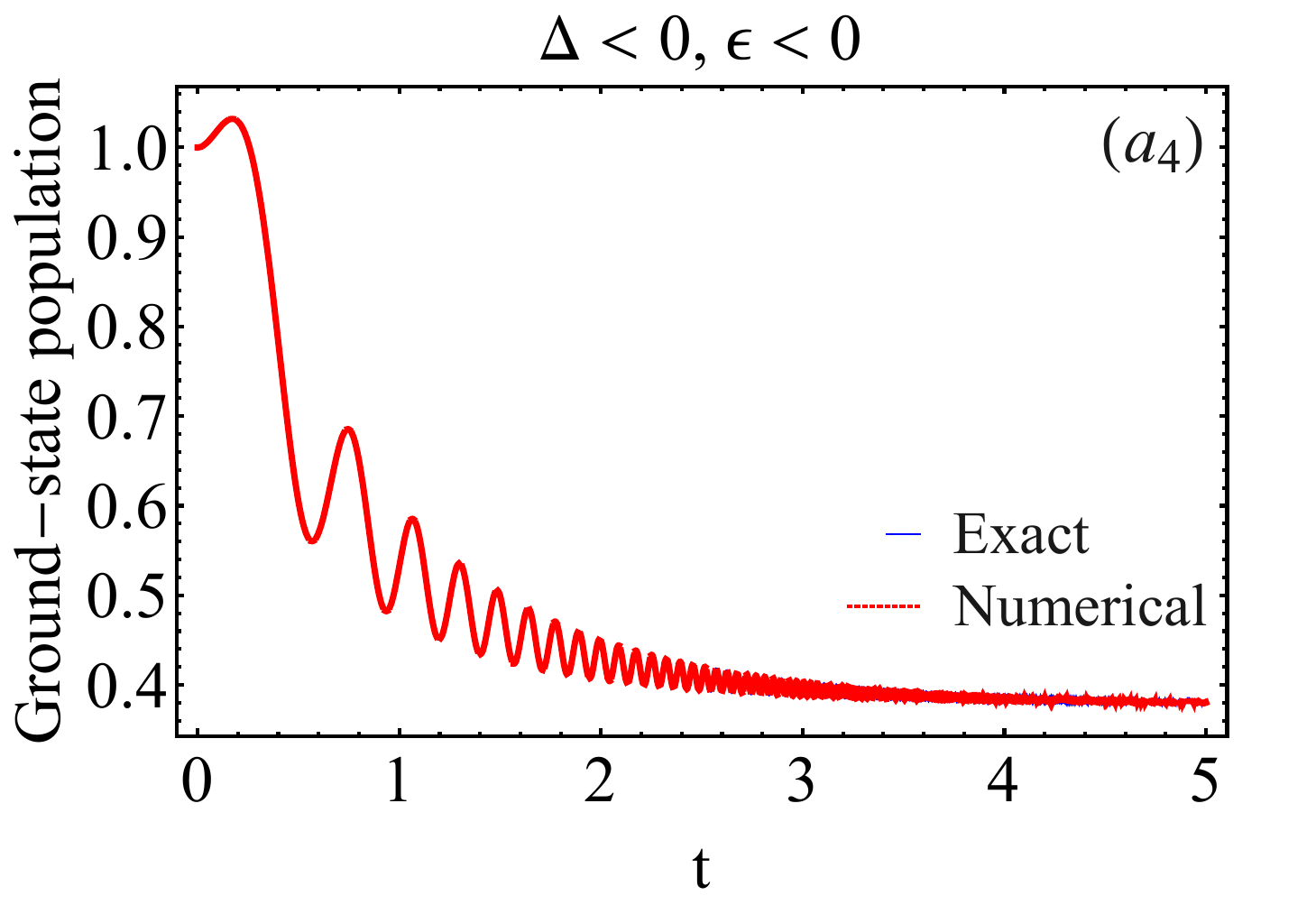}
	\includegraphics[width=0.22\textwidth, height =0.22\textwidth]{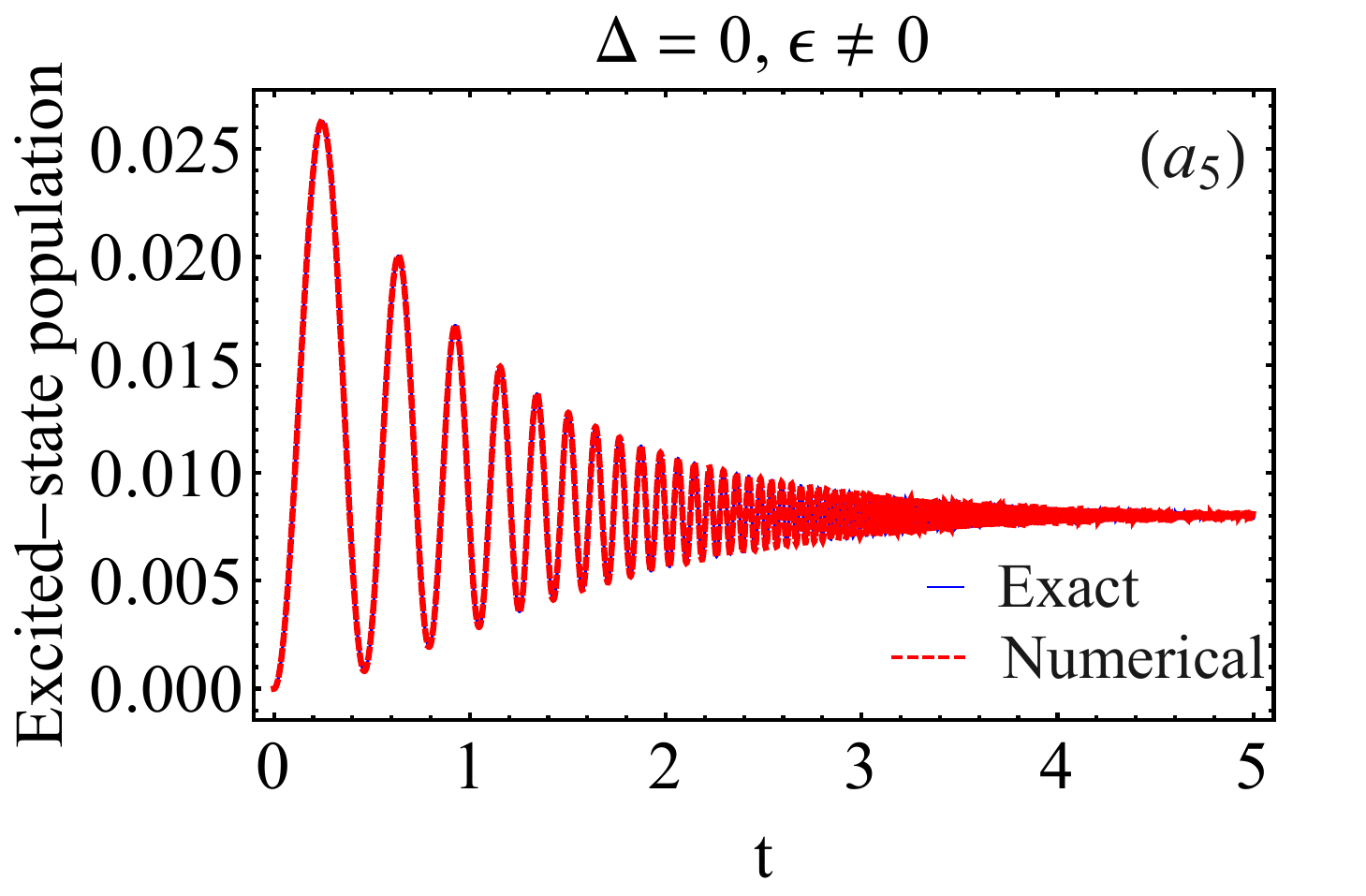}
	\includegraphics[width=0.22\textwidth, height =0.22\textwidth]{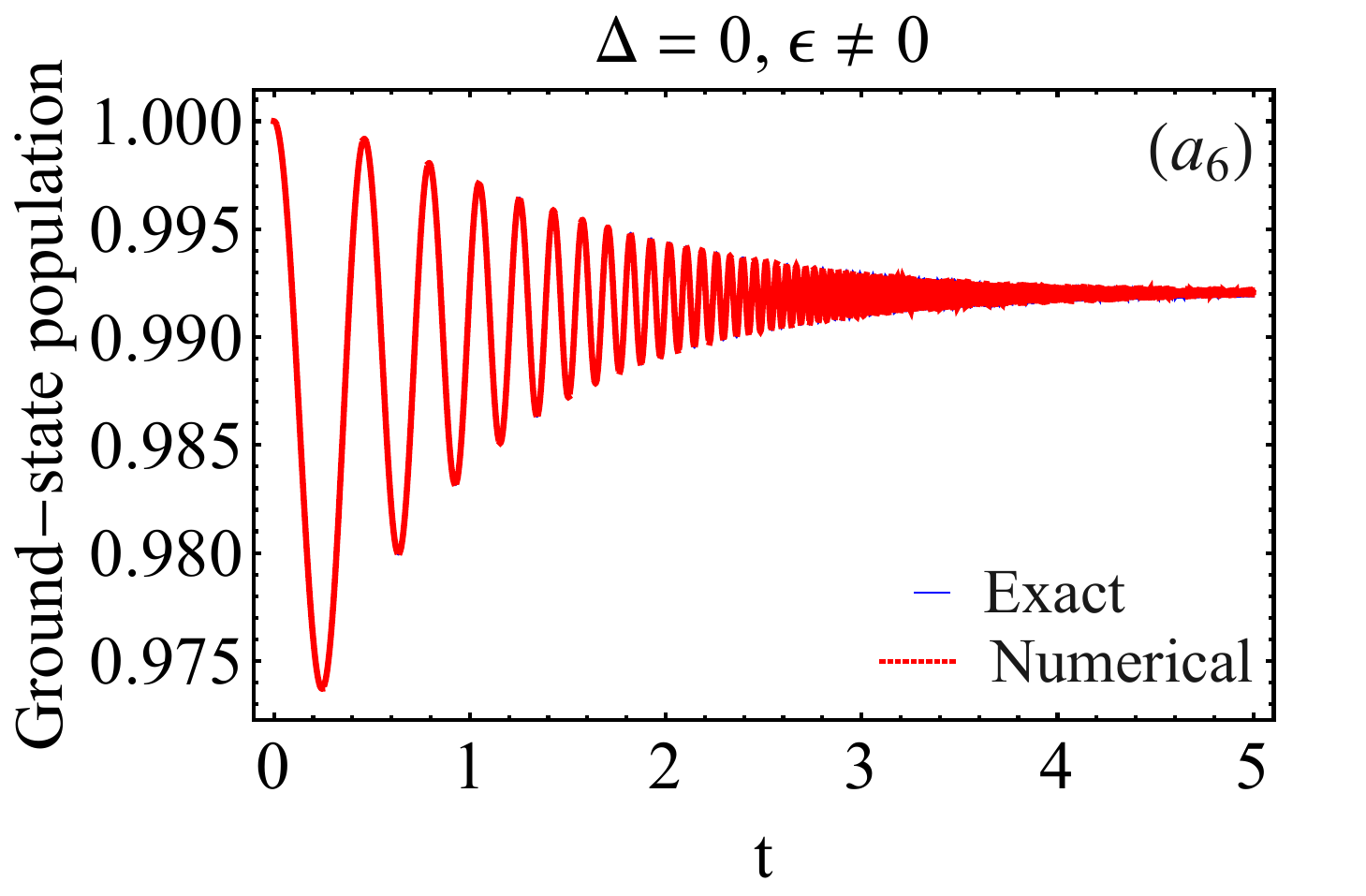}
	\caption{\small Variation of the different populations (ground and excited) of the first Nikitin model (Exp1) versus $\epsilon$. We have considered  $A= 2.0/\alpha,  \beta = 1.5/\alpha $. we remark that the shift and the imaginary coupling are responsible for the loss of information when their values are high in cases ($a_1$) and $(a_2)$. The exact results (Blue line) agree with the numerical solutions (red line).}\label{fig2}
\end{figure}

\begin{equation}
	{x^2}\frac{{{d^2}{\psi _1}\left( x \right)}}{{d{x^2}}} + x\left( {a - bx} \right)\frac{{d{\psi _1}\left( x \right)}}{{dx}} + {c^2}{\psi _1}\left( x \right) = 0.\label{2.3}
\end{equation}

As observed, the differential equation now depends on the variable $x$ incorporating the phase $\beta$, The parameters $a, b$ and $c$ are defined as follows:
\begin{equation}
	a = 1 + i\frac{\epsilon }{\alpha },\label{2.4}
\end{equation}
\begin{equation}
	b =  - i\frac{A}{\alpha },\label{2.5}
\end{equation}
\begin{equation}
	c = \frac{{i\Delta  + \epsilon }}{{2\alpha }}.\label{2.6}
\end{equation}

At this stage, these parameters acquire new roles: $a$ introduces a shift to the system, $b$ represents the signal strength, and $c$ captures the contribution of spontaneous emission, including the imaginary coupling. Since equation \eqref{2.3} is not directly solvable, we apply the ansatz transformation ${\psi _1}\left( x \right) = {x^\mu }z\left( x \right)$. This reformulates the differential equation into the confluent hypergeometric equation:

\begin{equation}
	{x^2}\frac{{{d^2}z\left( x \right)}}{{d{x^2}}} + \left( {\gamma  - bx} \right)\frac{{dz\left( x \right)}}{{dx}} - \mu bz\left( x \right) = 0,\label{2.7}
\end{equation}

\begin{figure}[h!]
	\centering
	\includegraphics[width=0.22\textwidth, height =0.22\textwidth]{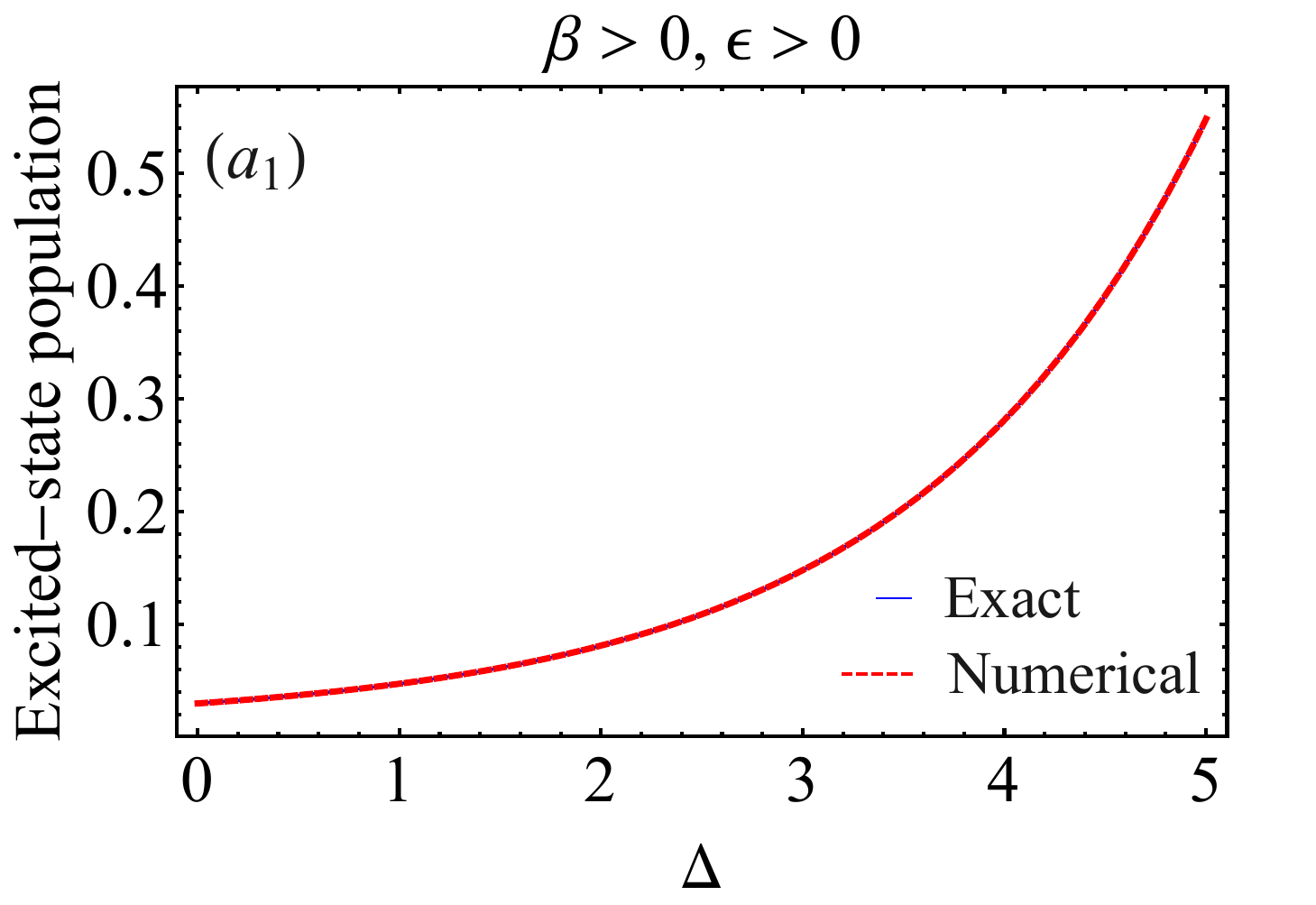}
	\includegraphics[width=0.22\textwidth, height =0.22\textwidth]{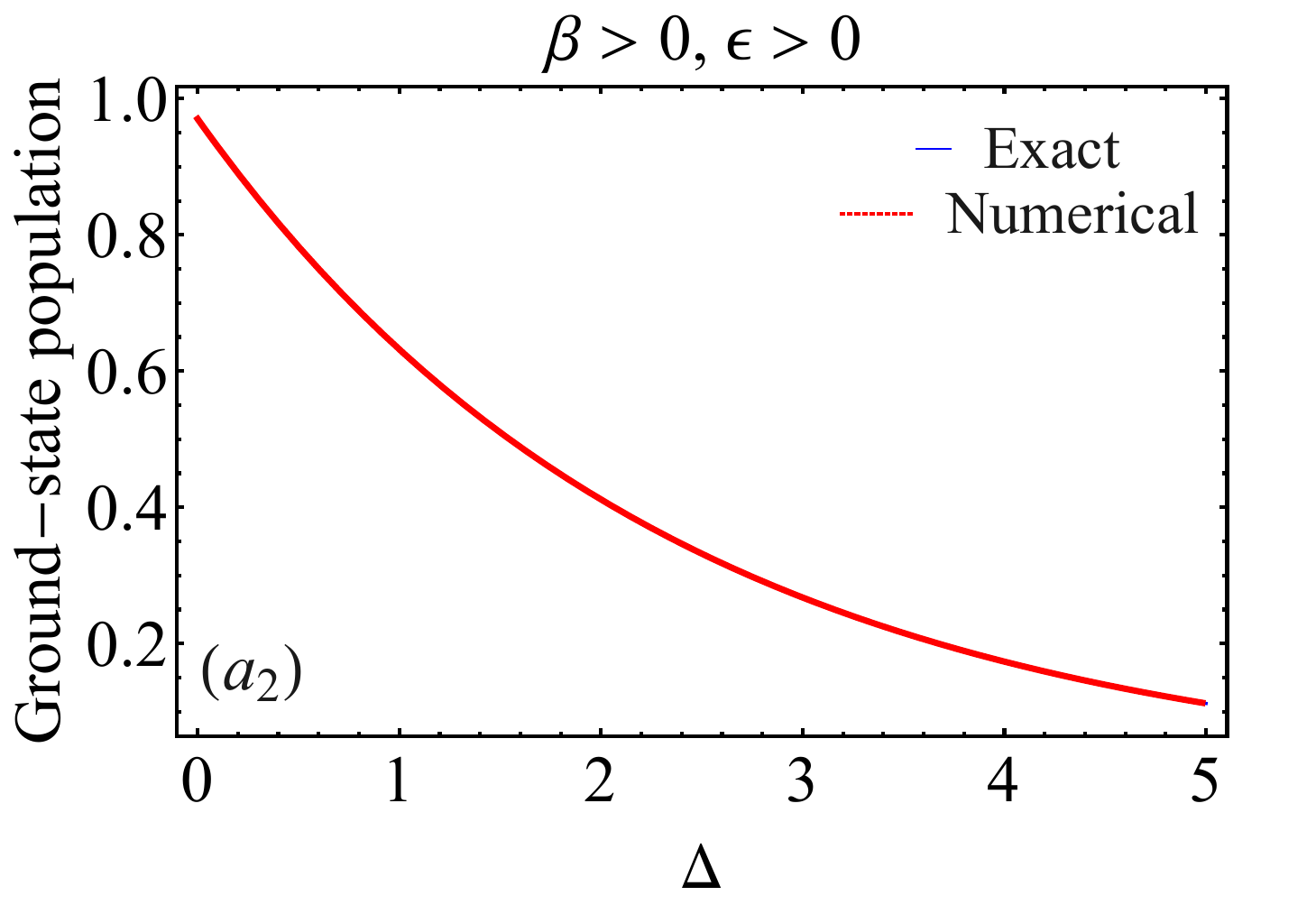}
	\includegraphics[width=0.22\textwidth, height =0.22\textwidth]{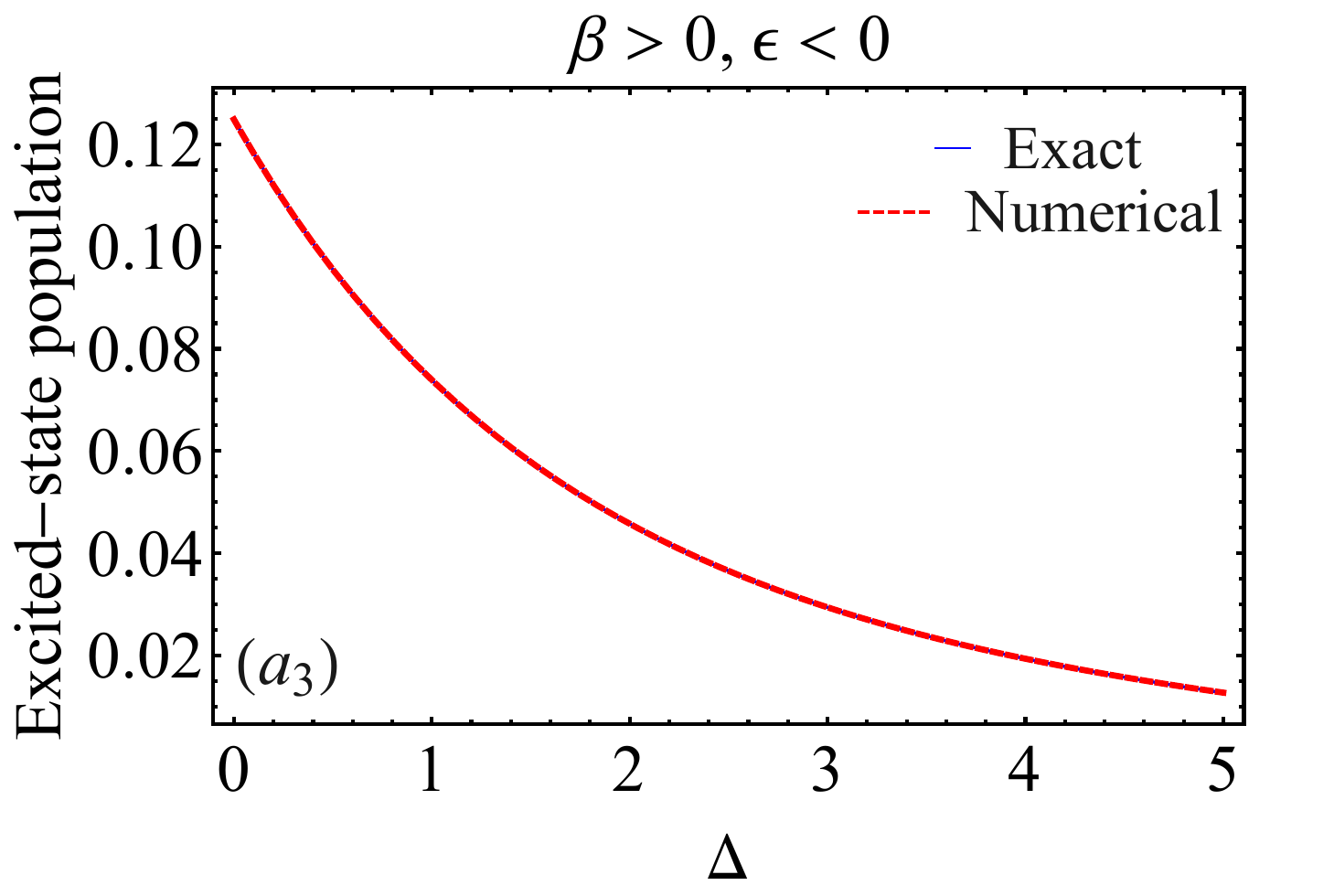}
	\includegraphics[width=0.22\textwidth, height =0.22\textwidth]{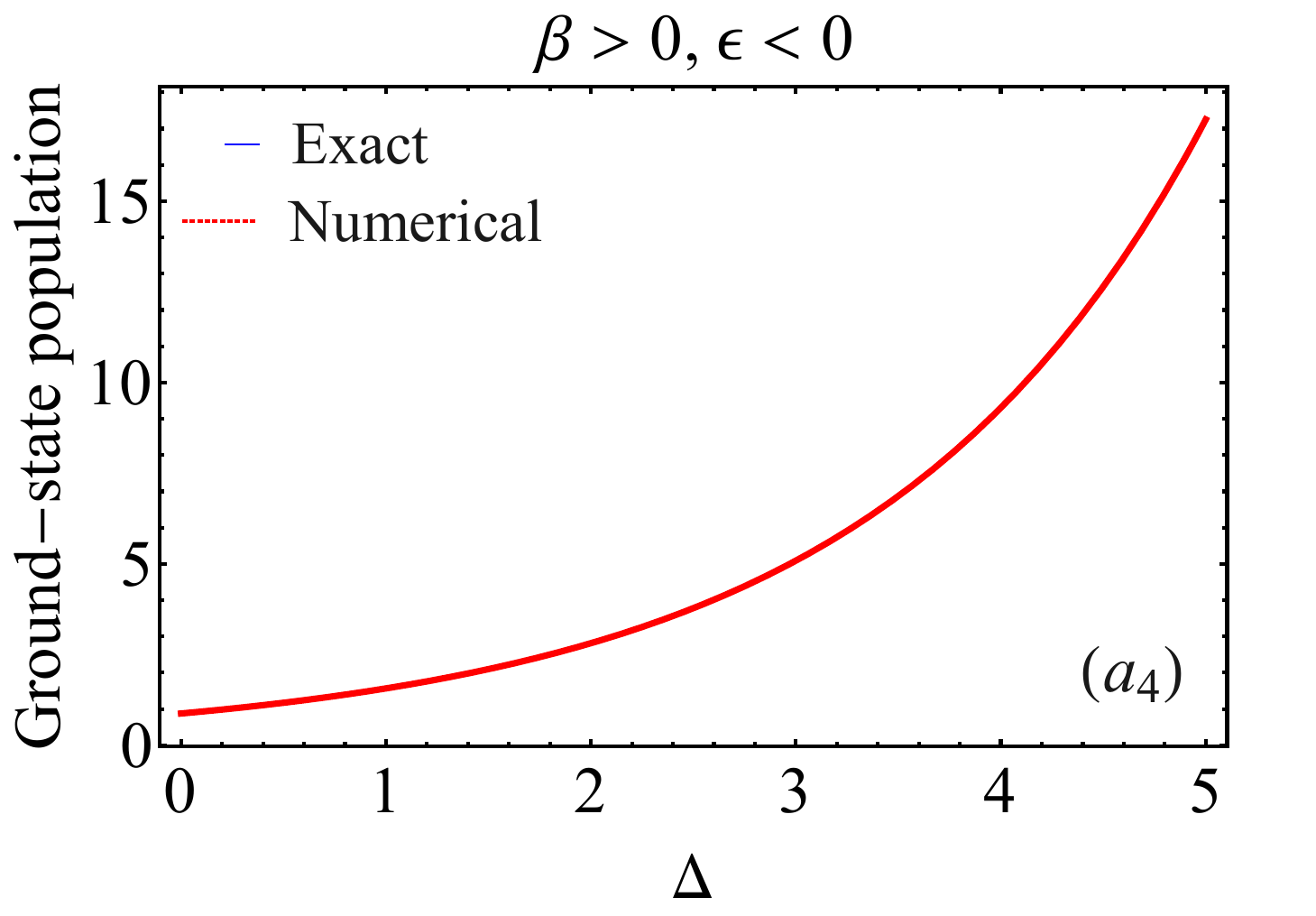}
	\includegraphics[width=0.22\textwidth, height =0.22\textwidth]{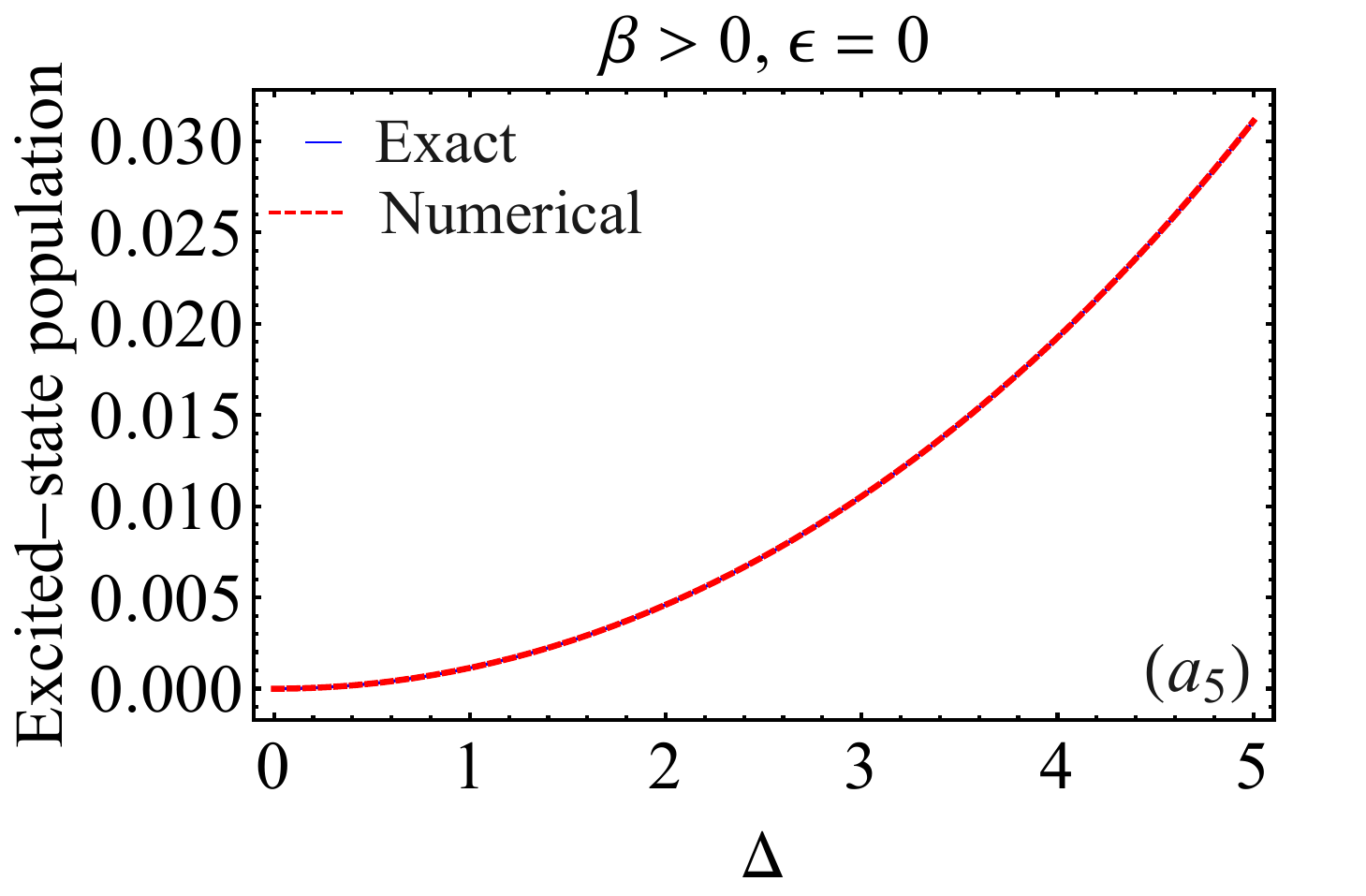}
	\includegraphics[width=0.22\textwidth, height =0.22\textwidth]{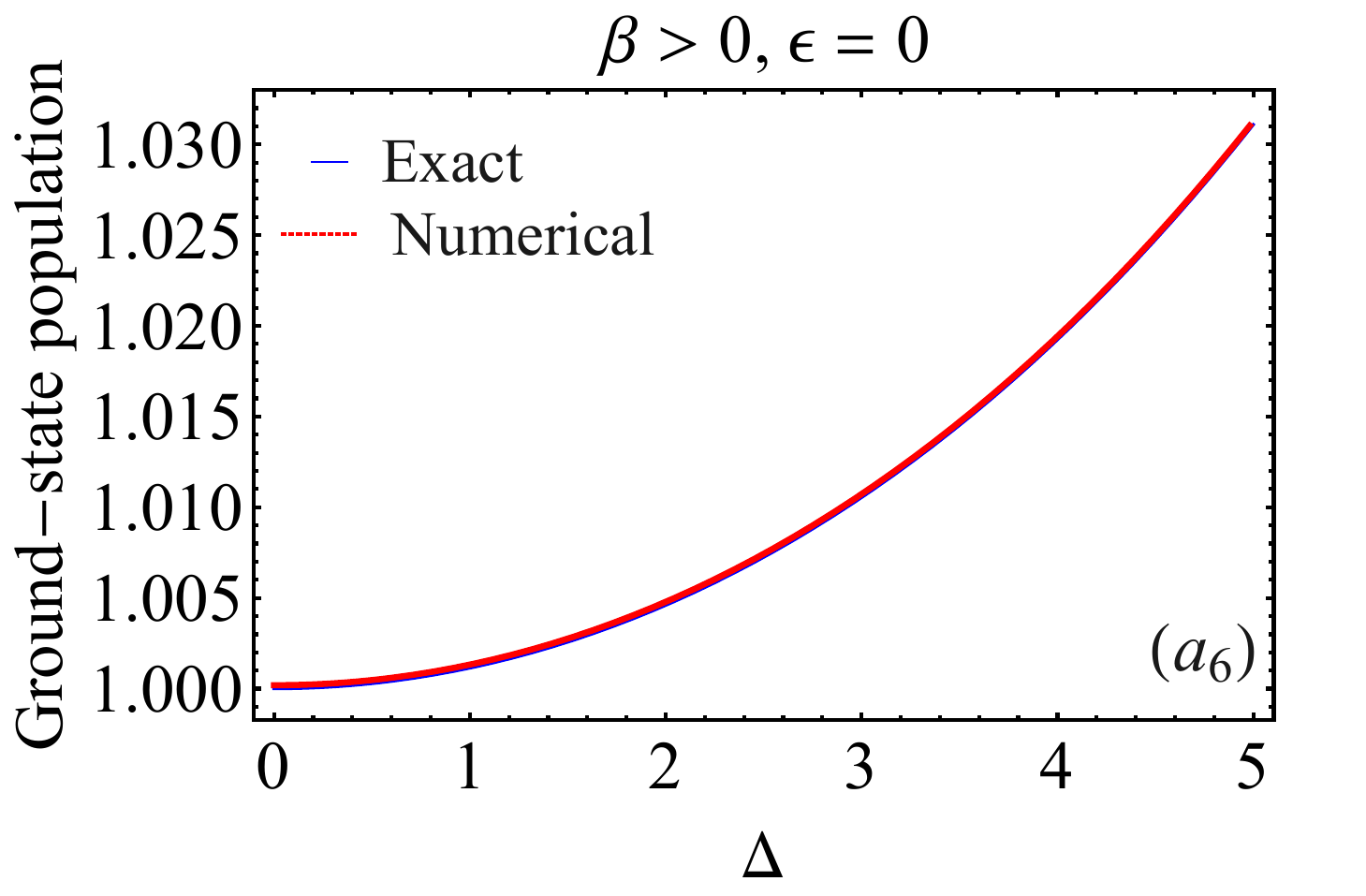}
	\caption{\small Variation of the different populations (ground state and excited state) of the first Nikitin model (Exp1) versus the coupling. We have considered  $A= 2.0,  t = 5,  \alpha = 1$. Exact results (Blue line)  agrees with the numerical solutions (red line).}\label{fig3}
\end{figure}

where the parameters $\gamma, b $, and $\mu$ are given by:
\begin{equation}
	{\mu _{1,2}} = \frac{{1 - a \mp \sqrt {{{\left( {1 - a} \right)}^2} - 4{c^2}} }}{2},\label{2.8}
\end{equation}

\begin{equation}
	\gamma  = 2\mu  + a. \label{2.9}
\end{equation}

\section{Results}

\subsection{Probabilities amplitudes}

The solution to equation \eqref{2.7} includes two linearly independent functions, $M\left( {\mu ,\gamma ,z} \right)$ and $U\left( {\mu ,\gamma ,z} \right)$.  Using the ansatz transformation, the probability amplitudes can be expressed as:
\begin{equation}
	{C_1}\left( {x,{x_0}} \right) = {x^\lambda }\exp \left( { - \frac{b}{2}x} \right)\left( {{a_ + }\left( {{x_0}} \right){U_1}\left( x \right) + {a_ - }\left( {{x_0}} \right){V_1}\left( x \right)} \right), \label{2.10}
\end{equation}

\begin{equation}
	{C_2}\left( {x,{x_0}} \right) = {x^\lambda }\exp \left( { - \frac{b}{2}x} \right)\left( {{a_ + }\left( {{x_0}} \right){U_2}\left( x \right) + {a_ - }\left( {{x_0}} \right){V_2}\left( x \right)} \right). \label{2.11}
\end{equation}

Here,

\begin{equation}
	{a_ \pm }\left( {{x_0}} \right) = {b_ \pm }x_0^{ - \lambda }\exp \left( {\frac{b}{2}{x_0}} \right),
\end{equation}
\begin{equation}
	{U_1}\left( x \right) = {x^\mu }M\left( {\mu ,\gamma ,z} \right),
\end{equation}
\begin{equation}
	{V_1}\left( x \right) = {x^\mu }U\left( {\mu ,\gamma ,z} \right),
\end{equation}
\begin{equation}
	{U_2}\left( x \right) =  - \frac{{i\mu }}{c}{x^\mu }\left( {M\left( {\mu ,\gamma ,z} \right) + \frac{x}{\gamma }M\left( {\mu  + 1,\gamma  + 1,z} \right)} \right),
\end{equation}
\begin{equation}
	{V_2}\left( x \right) =  - \frac{{i\mu }}{c}{x^\mu }\left( {U\left( {\mu ,\gamma ,z} \right) - xU\left( {\mu  + 1,\gamma  + 1,z} \right)} \right),
\end{equation}

The parameter $\lambda$ is defined as:

\begin{equation}
	\lambda  = \frac{{i\epsilon }}{{2\alpha }}.
\end{equation}

\begin{figure}[h!]
	\centering
	\includegraphics[width=0.22\textwidth, height =0.22\textwidth]{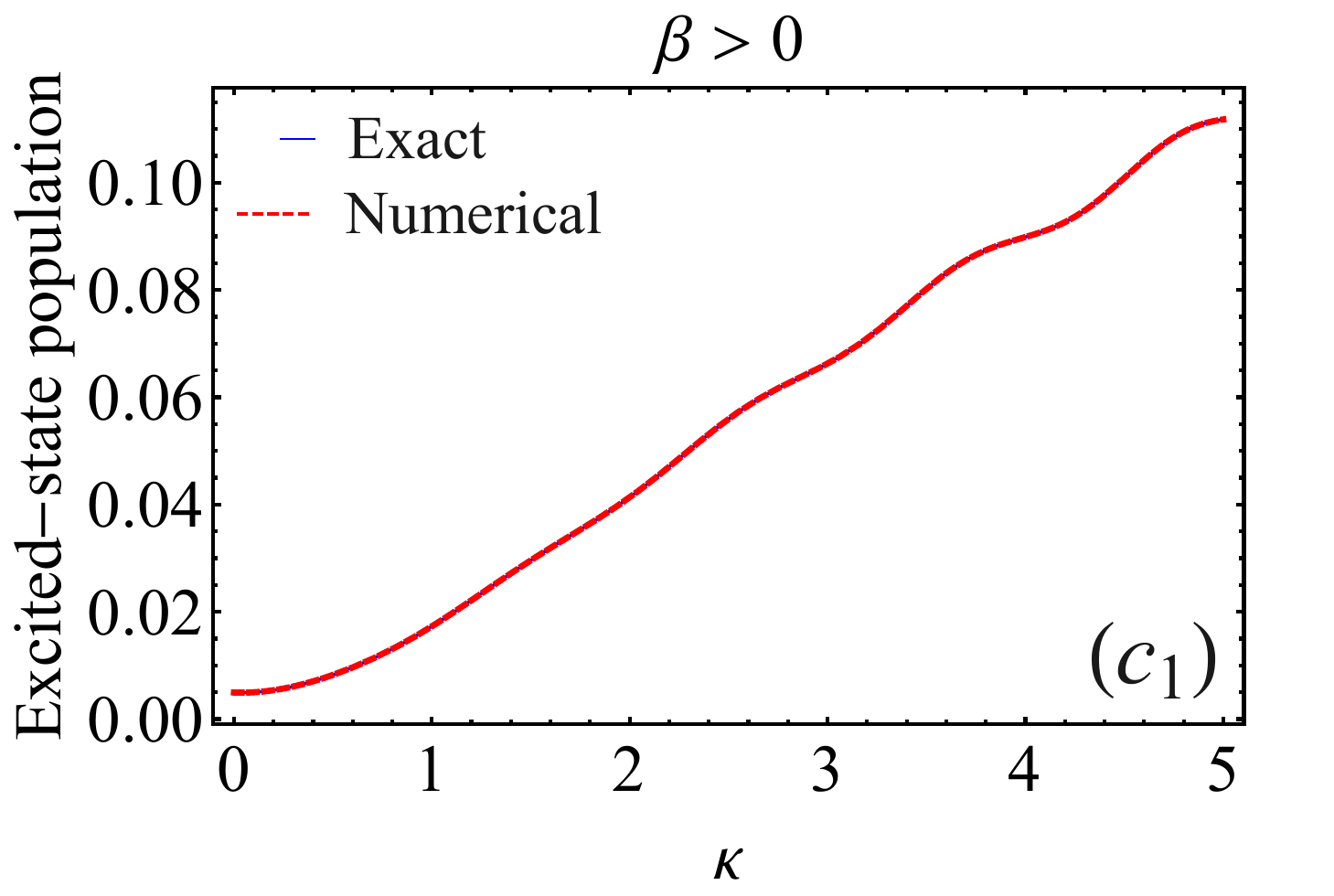}
	\includegraphics[width=0.22\textwidth, height =0.22\textwidth]{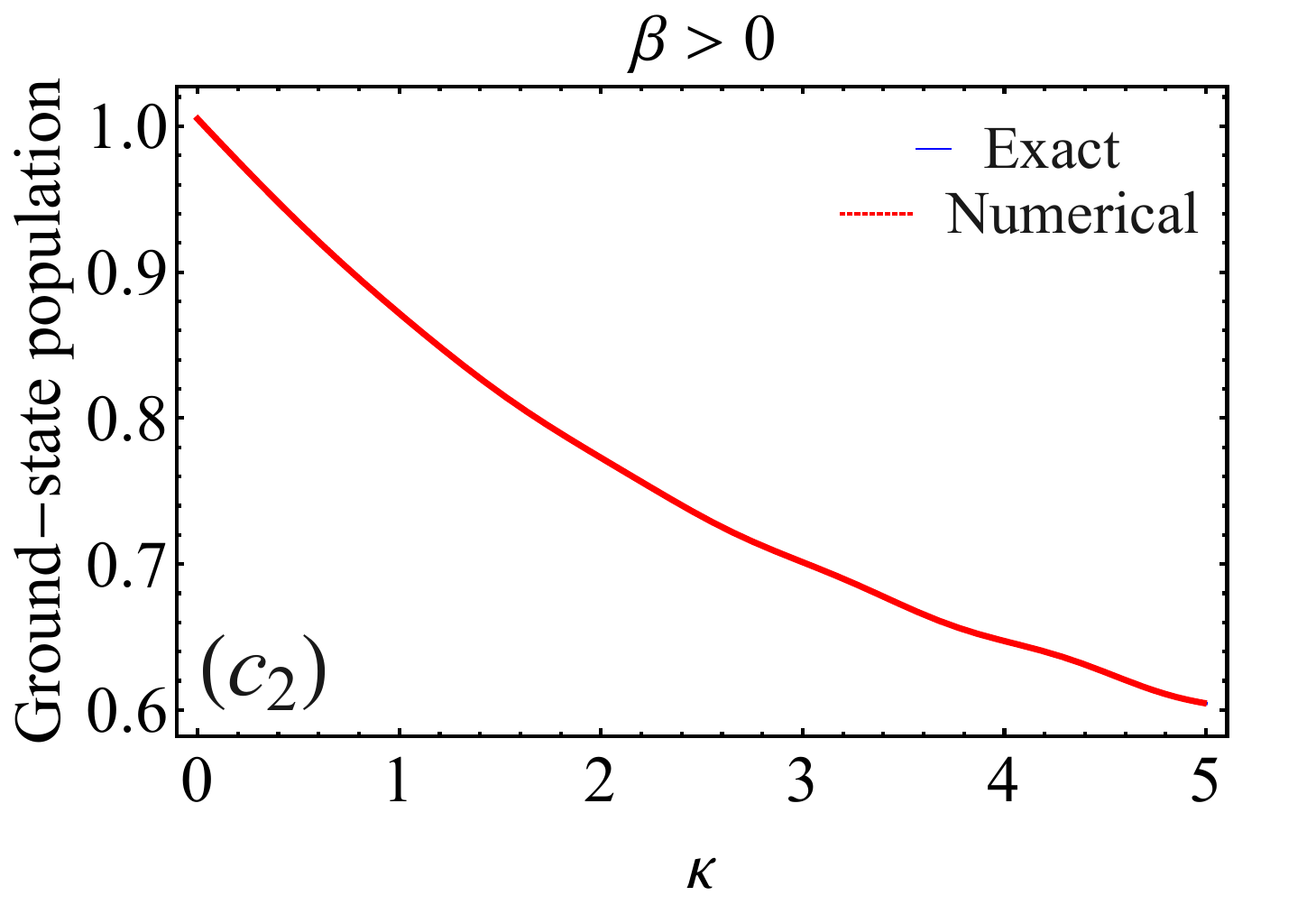}
	\includegraphics[width=0.22\textwidth, height =0.22\textwidth]{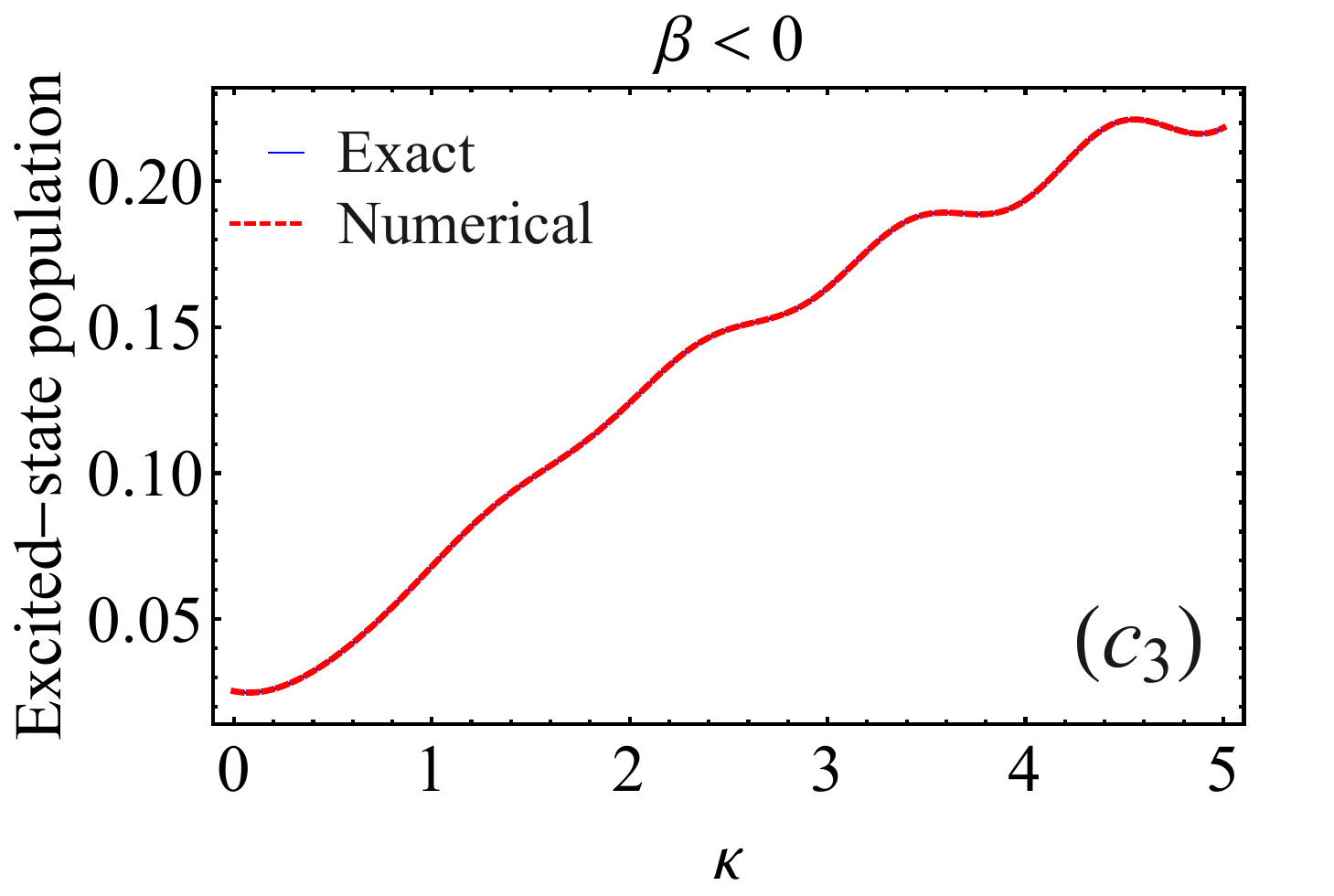}
	\includegraphics[width=0.22\textwidth, height =0.22\textwidth]{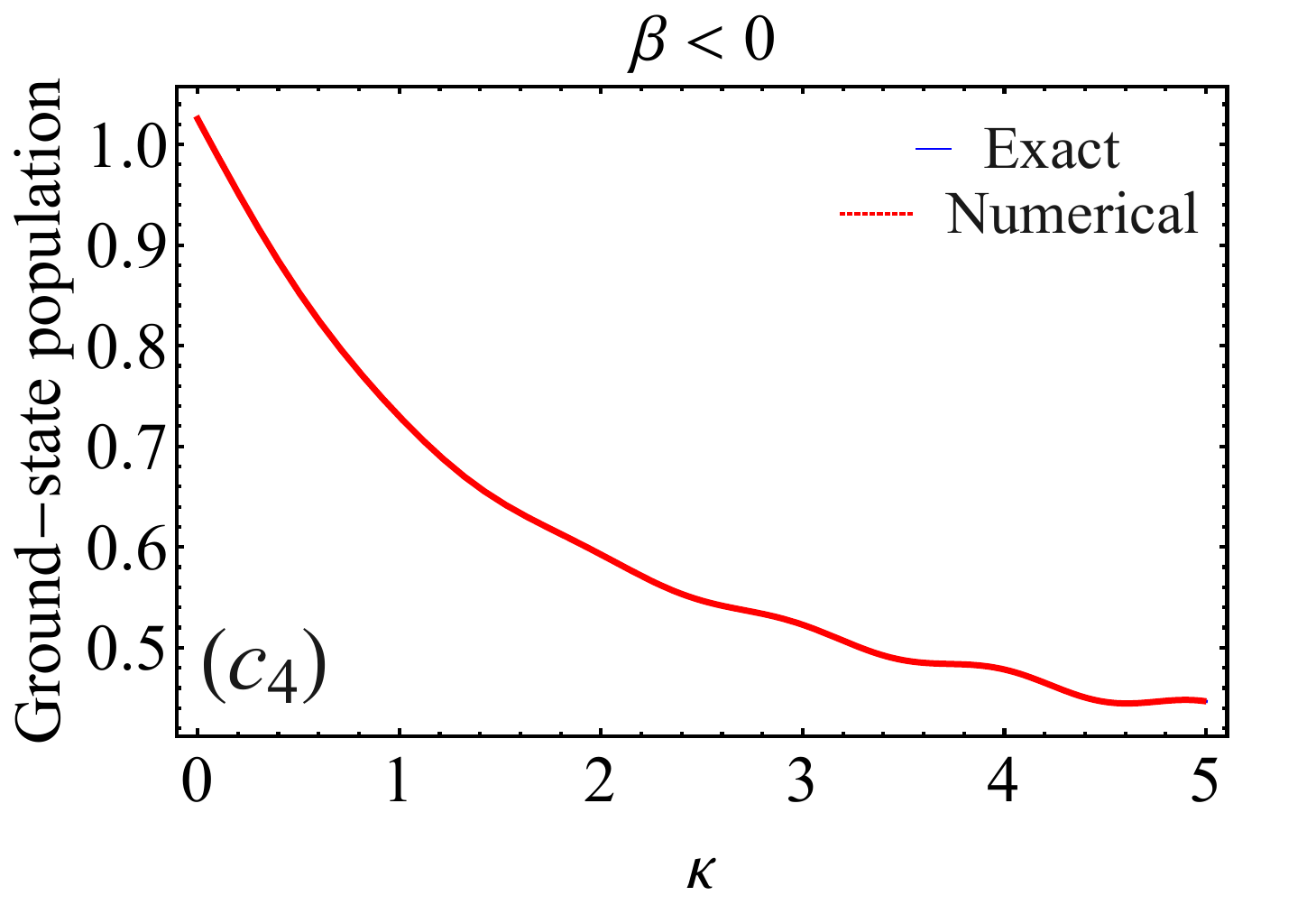}
	\includegraphics[width=0.22\textwidth, height =0.22\textwidth]{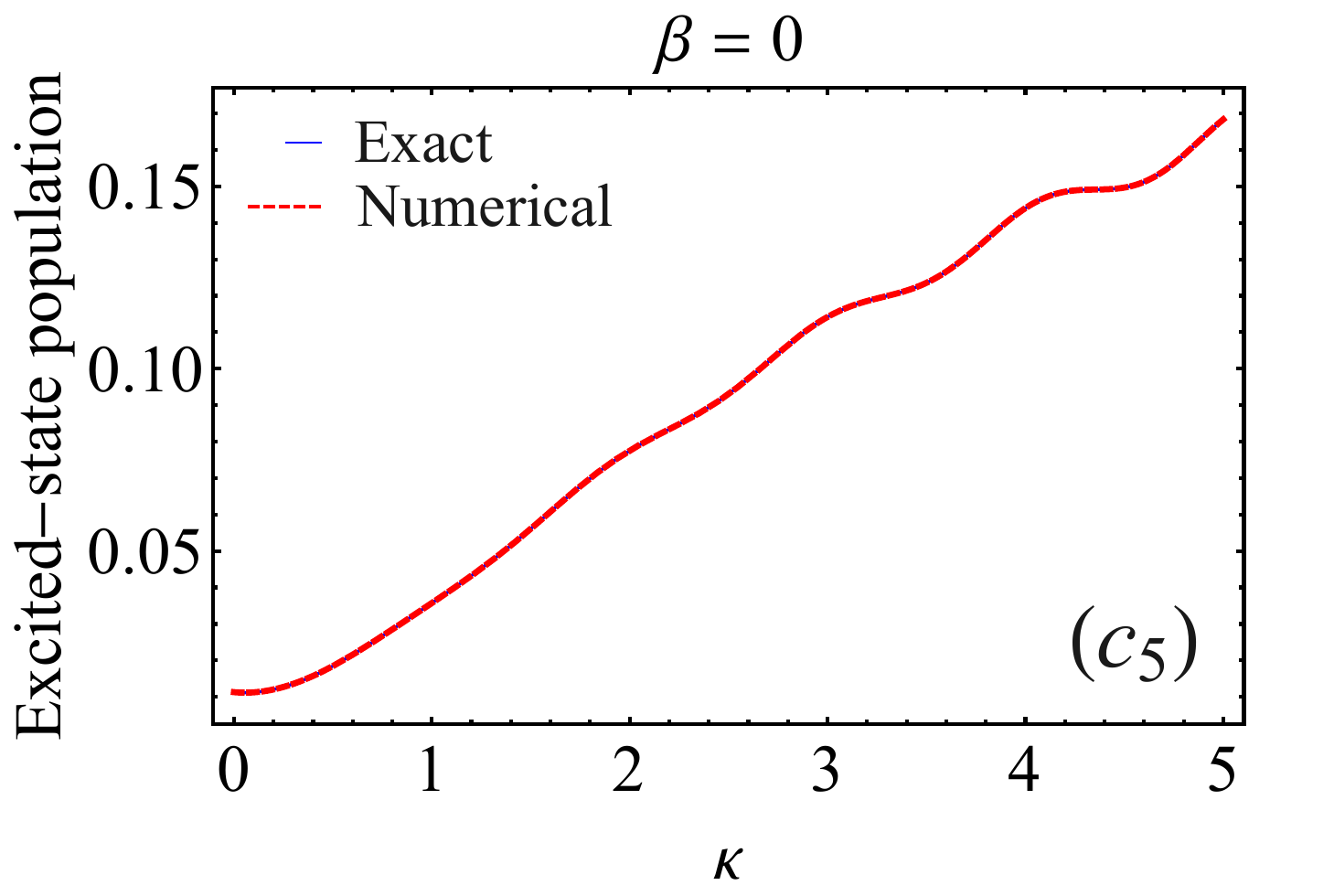}
	\includegraphics[width=0.22\textwidth, height =0.22\textwidth]{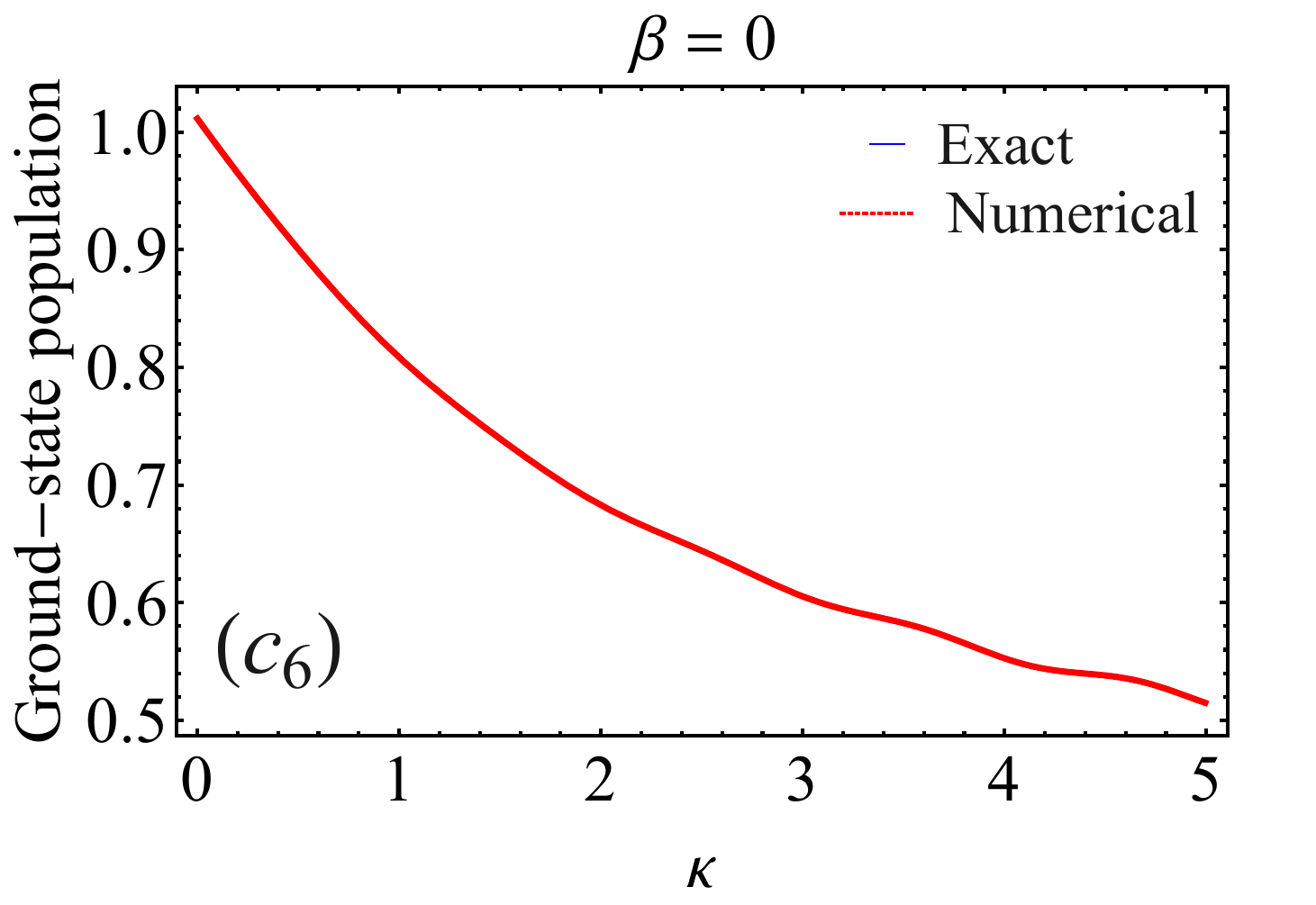}
	\caption{\small Evolution of the different populations at the ground state and excited state of the first Nikitin model (Exp1) versus the shift $\epsilon$ for different values of the decay rates. We have considered  $A= 2.0,  t = 5.0, \Delta = 0.5, \alpha = 1 $. The theoretical results (Blue line) agree with the numerical solutions (red line).}\label{fig4}
\end{figure}

The functions $M\left( {\mu ,\gamma ,x} \right)$ and $U\left( {\mu ,\gamma ,x} \right)$ represent Kummer's confluent hypergeometric functions. Their properties, as described in \cite{Bateman}, allow us to derive expressions for the derivatives:
\begin{equation}
	\frac{d}{{dz}}M\left( {\mu ,\gamma ,x} \right) = \frac{\mu }{\gamma }M\left( {\mu  + 1,\gamma  + 1,x} \right),
\end{equation}
\begin{equation}
	\frac{d}{{dz}}U\left( {\mu ,\gamma ,x} \right) =  - \mu U\left( {\mu  + 1,\gamma  + 1,x} \right).
\end{equation}

The constants ${b_ \pm }x_0^{ - \lambda }$ are determined using \cite{Kammogne1}.

\subsection{Propagator and probabilities}

To study the system's evolution between an initial time $t_0$ and an arbitrary time $t$, we use the propagator $C\left( x \right) = U\left( {x,{x_0}} \right)C\left( {{x_0}} \right)$. The components of the propagator are given by:
\begin{equation}
	{U_{k1}}\left( {x,{x_0}} \right) = \frac{{{\omega _{k1}}\left( {x,{x_0}} \right)}}{{{\omega _{12}}\left( {x,{x_0}} \right)}}\exp \left( { - \frac{b}{2}x} \right)\exp \left( {\frac{b}{2}{x_0}} \right),
\end{equation}
\begin{equation}
	{U_{k2}}\left( {x,{x_0}} \right) =  - \frac{{{\omega _{k2}}\left( {x,{x_0}} \right)}}{{{\omega _{12}}\left( {x,{x_0}} \right)}}\exp \left( { - \frac{b}{2}x} \right)\exp \left( {\frac{b}{2}{x_0}} \right).\label{2.21}
\end{equation}

The matrices elements are:
\begin{equation}
	{\omega _{kk'}}\left( {x,{x_0}} \right) = {V_k}\left( {{x_0}} \right){U_{k'}}\left( x \right) - {U_k}\left( {{x_0}} \right){V_{k'}}\left( x \right),
\end{equation}
\begin{equation}
	{\omega _{12}}\left( {{x_0},{x_0}} \right) = {U_2}\left( {{x_0}} \right){V_1}\left( {{x_0}} \right) - {U_1}\left( {{x_0}} \right){V_2}\left( {{x_0}} \right).
\end{equation}

Using the Wronskian of the confluent hypergeometric function:
\begin{equation}
	W = \frac{{\Gamma \left( \gamma  \right)}}{{\Gamma \left( \mu  \right)}}{x^{ - \gamma }}\exp \left( x \right),\label{2.22}
\end{equation}

we calculate the transition parameter between levels $1$ to level $2$
\begin{equation}
	{\omega _{12}}\left( {{x_0},{x_0}} \right) = \frac{i}{c}\frac{{\Gamma \left( \gamma  \right)}}{{\Gamma \left( \mu  \right)}}x_0^{2\mu  - \gamma  + 1}\exp \left( {{x_0}} \right),
\end{equation}

responsible for the transition between the ground and the excited state. The propagator element is expressed by
\begin{equation}
	{U_{12}}\left( {x,{x_0}} \right) = ic\frac{{\Gamma \left( \mu  \right)}}{{\Gamma \left( \gamma  \right)}}\left( {F\left( {x,{x_0}} \right) - F\left( {{x_0},x} \right)} \right)\exp \left( {\vartheta \left( {x,{x_0}} \right)} \right),
\end{equation}
\begin{equation}
	\begin{gathered}
		{U_{22}}\left( {x,{x_0}} \right) = \mu {b^{\gamma  - 1}}\frac{{\Gamma \left( \mu  \right)}}{{\Gamma \left( \gamma  \right)}}(F\left( {x,{x_0}} \right) - F\left( {{x_0},x} \right) \hfill \\
		+ \frac{{bx}}{\gamma }M\left( {\mu  + 1,\gamma  + 1,{x_0}} \right)U\left( {\mu ,\gamma ,x} \right) \hfill \\
		+ b{x_0}M\left( {\mu  + 1,\gamma  + 1,x} \right)U\left( {\mu ,\gamma ,{x_0}} \right))\exp \left( {\vartheta \left( {x,{x_0}} \right)}. \right) \hfill \\ 
	\end{gathered}
\end{equation}

The two matrices of the propagator are necessary to describe the transition or the survival of the particle. we easily determine the survival probability given by:
\begin{equation}
	{P_{22}}\left( {x,{x_0}} \right) = \operatorname{Re} \left( {{U_{22}}\left( {x,{x_0}} \right)} \right) + \operatorname{Im} \left( {{U_{22}}\left( {x,{x_0}} \right)} \right).
\end{equation}

The probability of transition is given by:
\begin{equation}
	{P_{12}}\left( {x,{x_0}} \right) = \operatorname{Re} \left( {{U_{12}}\left( {x,{x_0}} \right)} \right) + \operatorname{Im} \left( {{U_{12}}\left( {x,{x_0}} \right)} \right).
\end{equation}

\subsection{Interpretation of the results}

Figures \ref{fig2}, \ref{fig3}, and \ref{fig4}, describe the evolution of the ground and excited-state populations during spontaneous emission.

In Fig. \ref{fig2}, we observe the evolution of the first Nikitin model for the ground and excited-state populations as a function of time when the coupling and the shift vary. Regarding the excited-state population, when $\Delta > 0$ and $\epsilon > 0$  the system demonstrates an increase in oscillations with a weak amplitude, as shown in panel $(a_1)$. Conversely, when $\Delta < 0$ and $\epsilon < 0$, as depicted in panel $(a_3)$, the system also oscillates, but its amplitude doubles. This indicates that the imaginary coupling and the shift enhance information transfer at low values. When $\Delta = 0$ and $\epsilon \ne 0$ in panel $(a_5)$, oscillations occur with very low amplitude due to the absence of coupling. In the ground-state population, oscillations decrease whether the imaginary coupling increases $(a_2)$ or decreases $(a_4)$,  implying that this parameter does not significantly influence population evolution. However, when $\Delta = 0$ and $\epsilon \ne 0$, the oscillations increase, and transmission is maximal when the imaginary part of the coupling vanishes.

In Fig. \ref{fig3}, we plot the evolution of the different populations against the coupling when the phase $\beta$ and the shift $\epsilon$ vary. In both the ground and excited states, the two populations exhibit exponential decay due to the absence of oscillations. This decay is less pronounced in panels $(a_1), (a_2), (a_3), (a_5)$. However, in panels $(a_4) and (a_6)$, the decay is more pronounced due to the high values of the shift in the ground-state population. In this case, the population exceeds one, an effect attributed to the high phase. Therefore, the shift and phase are responsible for the rapid decay of the population in this scenario.

Figure \ref{fig4} illustrates the variation of the previous states versus the shift $\epsilon$ as the phase varies. In panels $(c_2), (c_4)$ and $(c_6)$ the population decreases exponentially. In contrast, panels $(c_1), (c_3)$ and $(c_5)$ exhibit slight oscillations due to the high value of the phase $\beta$ in the excited-state population. This implies that the phase $\beta$ is critical for the potential transmission of information in the ground-state population.

\section{Energies of the Nikitin model}

To study the energies of the Nikitin model, we analyze the eigenvalues of our system, which yield:
\begin{equation}
	{E_ \pm } =  \pm \left( {i\Delta  + \epsilon } \right)\csc \left( {2\vartheta \left( t \right)} \right),
\end{equation}

\begin{figure}[h!]
	\centering
	\includegraphics[width=0.22\textwidth, height =0.22\textwidth]{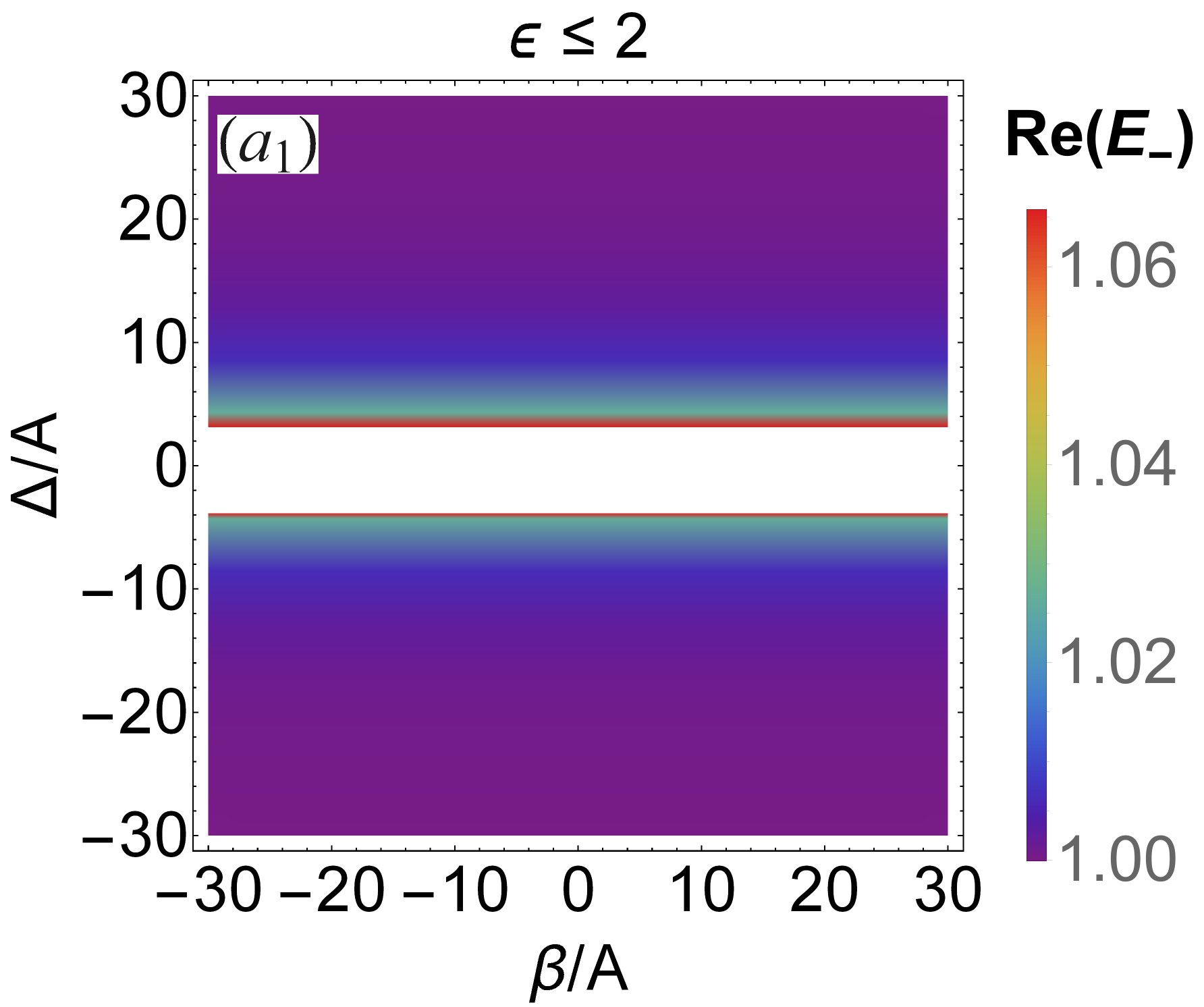}
	\includegraphics[width=0.22\textwidth, height =0.22\textwidth]{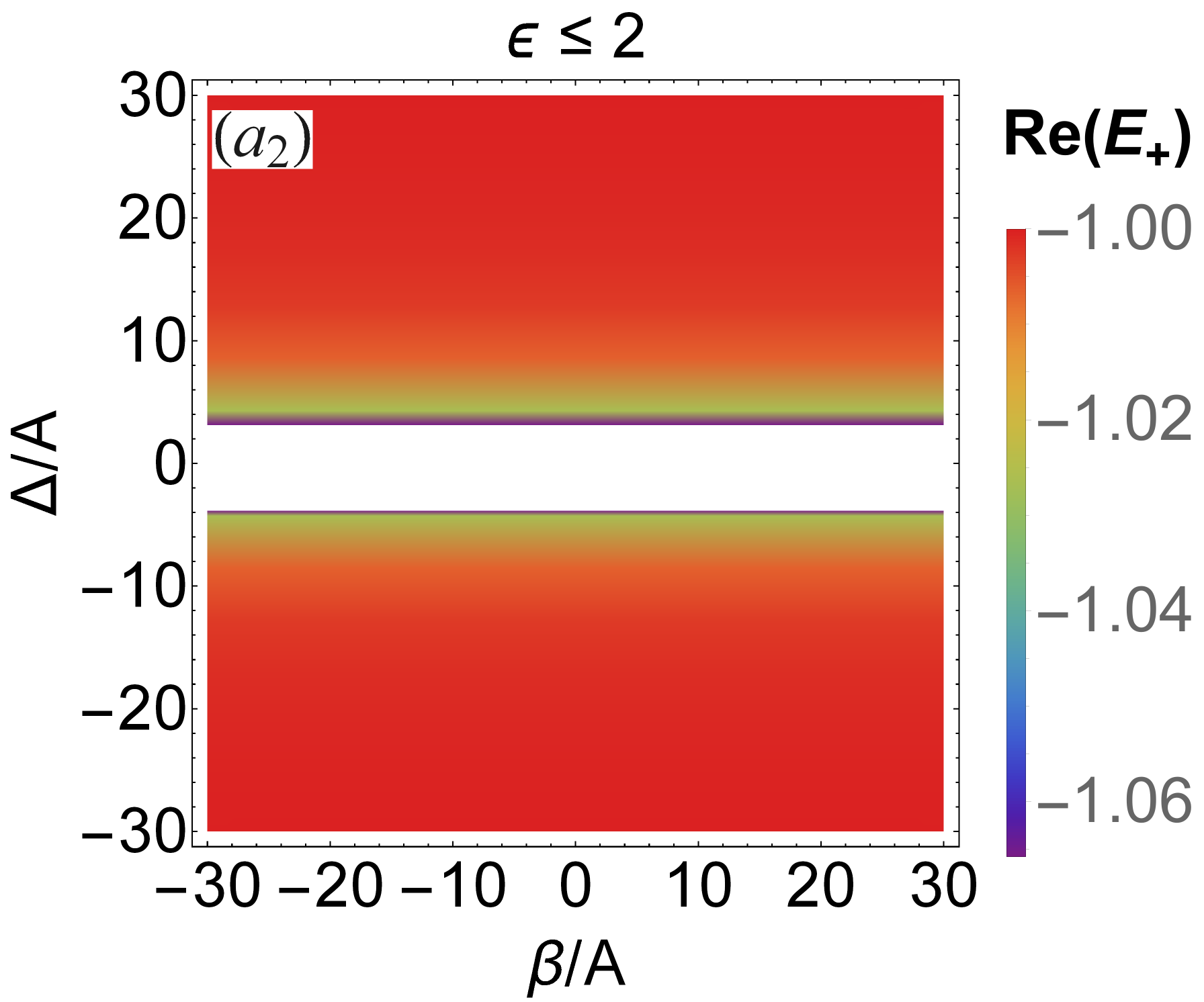}
	\includegraphics[width=0.22\textwidth, height =0.22\textwidth]{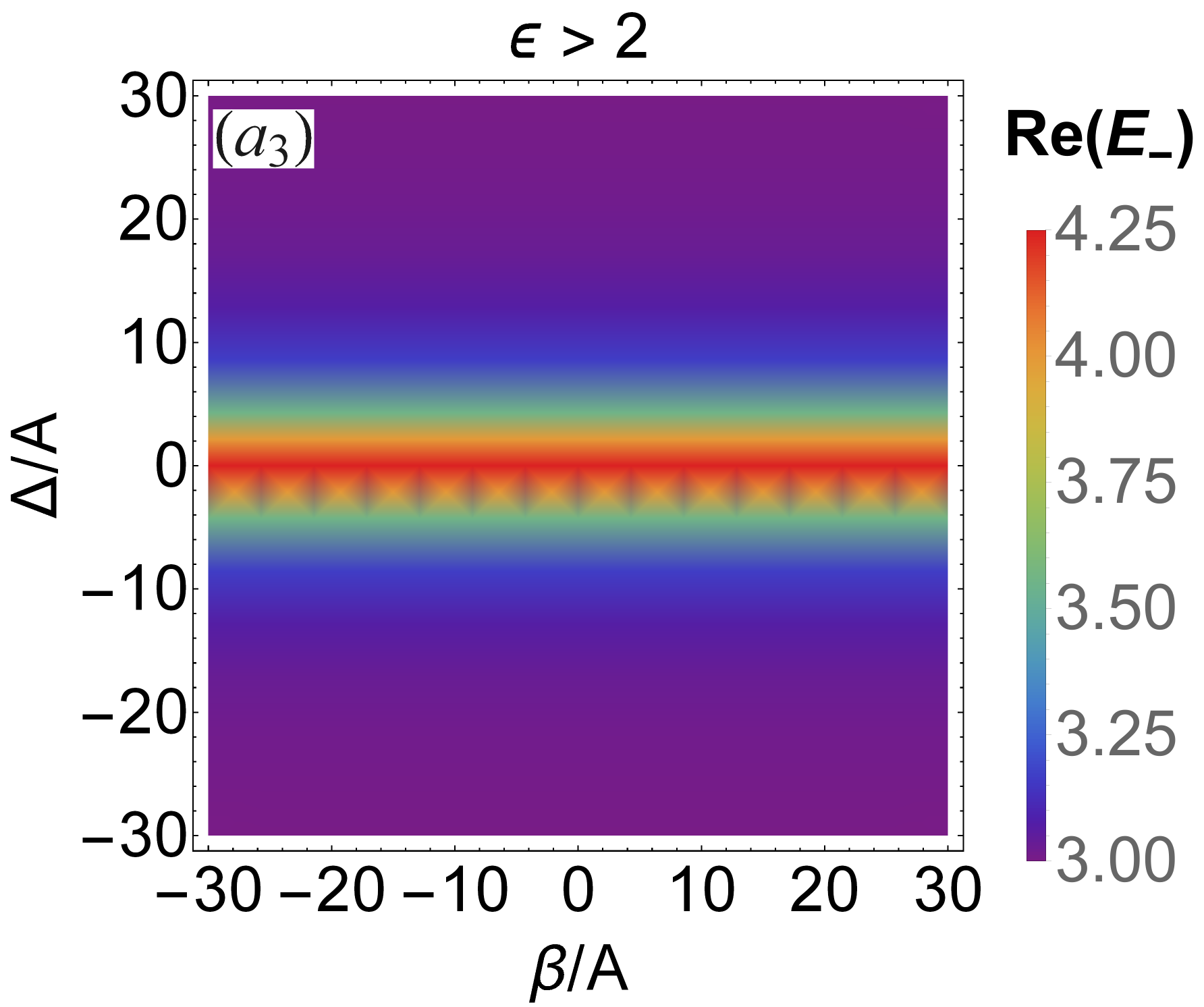}
	\includegraphics[width=0.22\textwidth, height =0.22\textwidth]{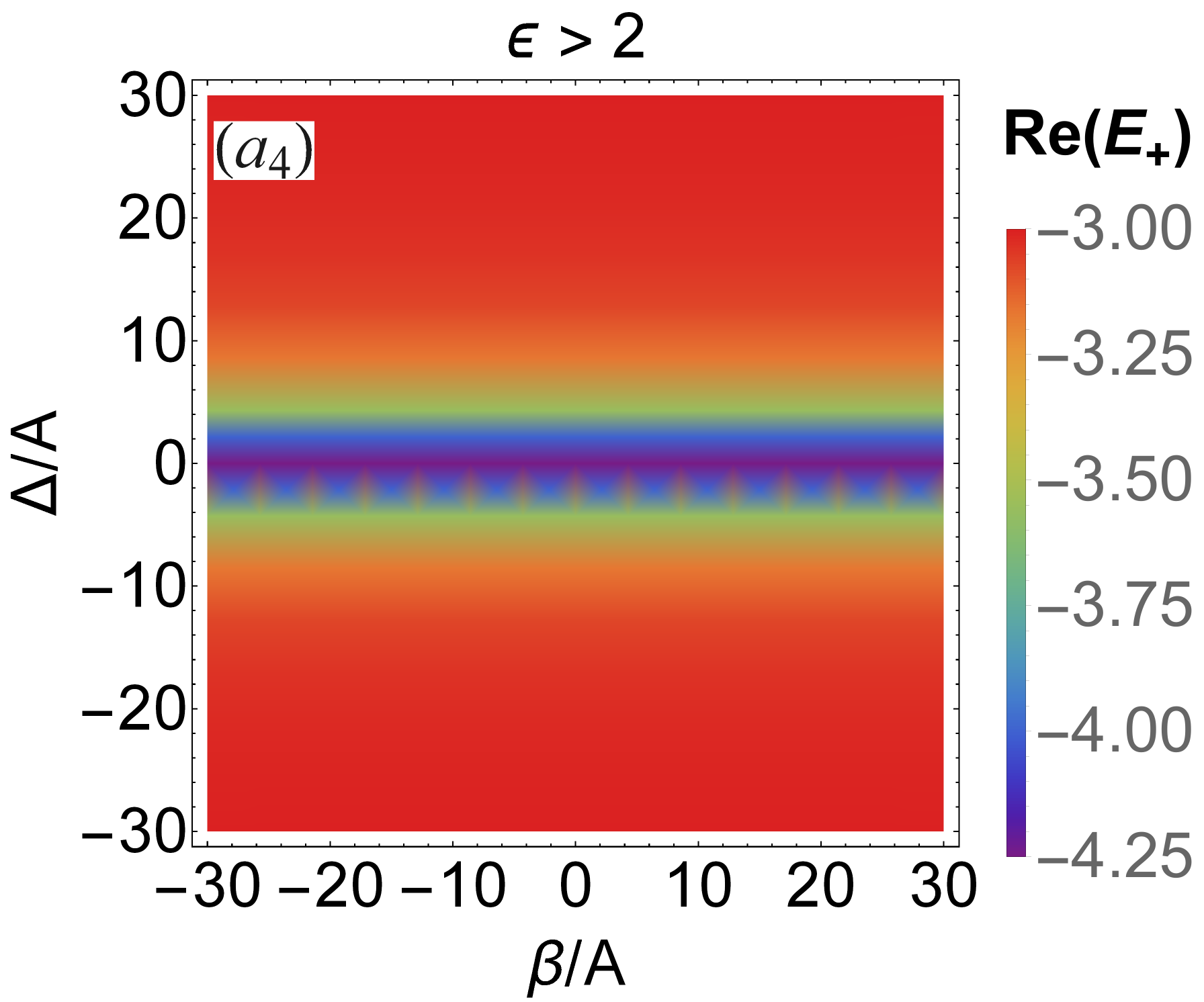}
	\caption{\small Permitted and forbidden zone on the energy diagram of the real part. We have considered  $\alpha= -15.0/A,  t = 7.0/A $. We use Mathematica to plot these figures and we choose the Rainbow color as color function. The shift plays a great function in the creation of the barrier to control quantum information.}\label{fig5}
\end{figure}

where
\begin{equation}
	\tan \left( {2\vartheta \left( t \right)} \right) = - \frac{{i\Delta  + \epsilon }}{{\Omega \left( t \right)}}.
\end{equation}

The eigenenergy is complex due to the imaginary coupling term, which highlights the non-Hermitian nature of this model. The eigenenergy can be separated into real and imaginary parts:

\begin{equation}
	\operatorname{Re} \left( E_\pm \right) =  \frac{1}{2}\left| {{Z^{1/2}}\left( t \right)} \right|\cos \phi,
\end{equation}
\begin{equation}
	\operatorname{Im} \left( E_\pm \right) = \frac{1}{2}\left| {{Z^{1/2}}\left( t \right)} \right|sin\phi,
\end{equation}

\begin{figure}[h!]
	\centering
	\includegraphics[width=0.23\textwidth, height =0.20\textwidth]{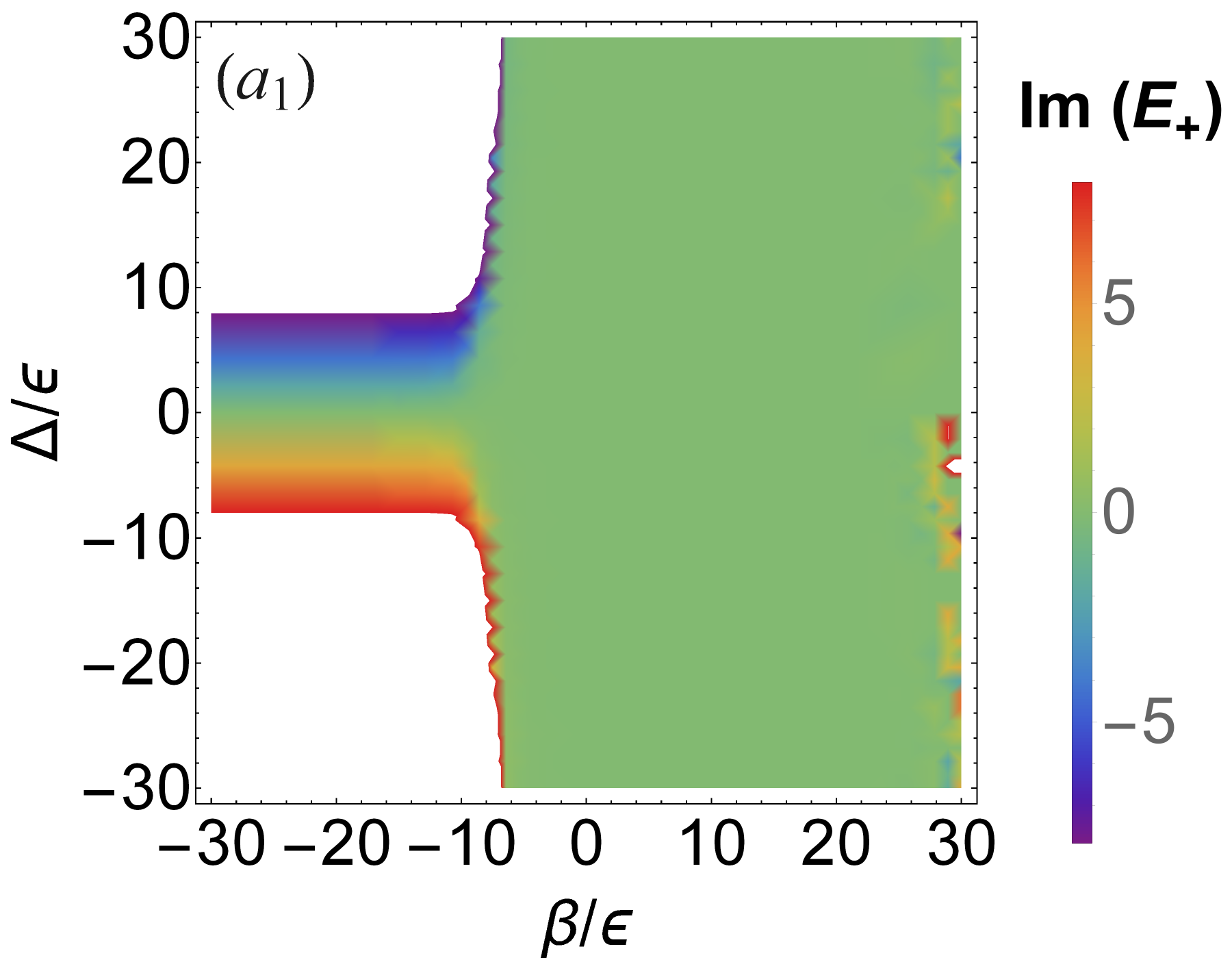}
	\includegraphics[width=0.23\textwidth, height =0.20\textwidth]{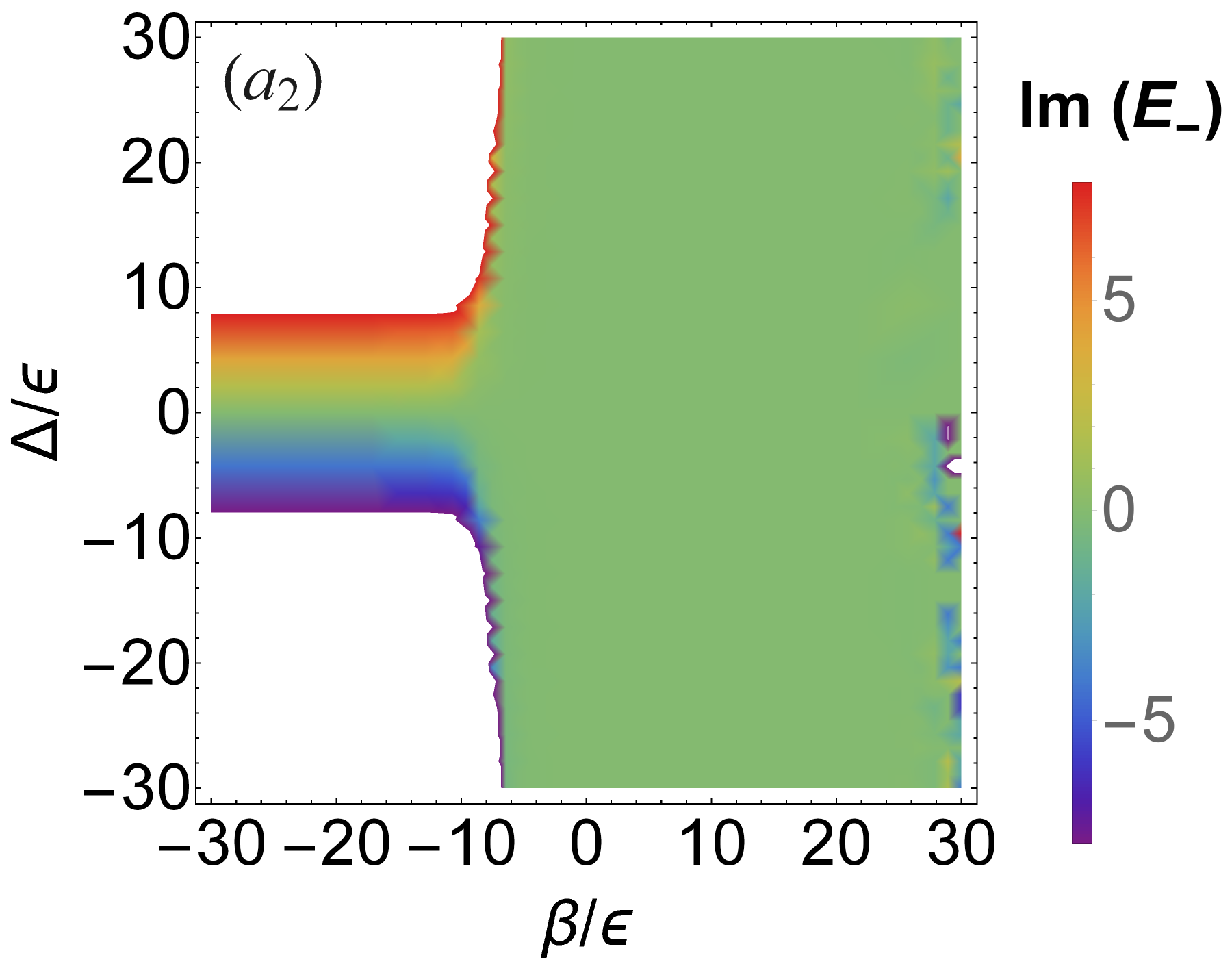}
	\caption{\small (Color online) The imaginary part of the energy against the coupling and the phase. In this figure, we have considered  $A/= 20/\epsilon,  t = 15/\epsilon, \alpha = 0.5/\epsilon $. The sweep velocity $\alpha$ and the time $t$ are responsible for the order or the chaos in the system for high values. Remind that the imaginary part of the energy describes the loss of quantum information.}\label{fig6}
\end{figure}

with
\begin{equation}
	\tan \left( {2\phi } \right) = \frac{{2\epsilon \Delta }}{{{\Omega ^2}\left( t \right) + {\epsilon ^2} - {\Delta ^2}\left( t \right)}},
\end{equation}
\begin{equation}
	{Z^{1/2}}\left( t \right) = \frac{{2\epsilon \Delta }}{{{\Omega ^2}\left( t \right) + {\epsilon ^2} - {\Delta ^2}\left( t \right)}}.
\end{equation}

Figures \ref{fig5} and \ref{fig6} provide graphical representations of the real and imaginary parts of the energy. The real part indicates energy gain, while the imaginary part reflects information loss.

In Fig.\ref{fig5}, the color bar $Re(E_-))$ and $Re(E_+))$ illustrates the evolution of eigenenergies for the ground and excited states. We observe a transmission or forbidden zone for specific information. When $\epsilon \leqslant 2$, the system creates a barrier (white color) that prevents information transmission near zero on the coupling axis (panels $(a_1)$ and $(a_2)$). When $\epsilon > 2$, the system generates a barrier (blue color) that allows information transmission in the same zone. Both phenomena occur over a long phase duration.

Figure \ref{fig6} shows the imaginary part of the energy as a function of the phase and coupling. The color bar depicts the variation in imaginary coupling for the ground $(a_1)$ and excited states $(a_2)$. The transmission of information is localized (between $-30/\alpha$ and $-10/\alpha$) due to information loss. The sweep velocity and amplitude influence order or chaos in the system for high values. The imaginary part describes the system's quantum information loss.

\section{Similarity with the Rabi model}

The first exponential Nikitin model shows similarities with the Rabi model when time approaches $- \infty$ in an exponential function. Under these conditions, the detuning becomes $\Omega \left( t \right) = \epsilon $, while the coupling remains a complex function $i\Delta  + \epsilon $. To study the dynamics of this model, we again use the time-dependent Schr\"odinger equation. After applying appropriate gauge transformations, the differential equation for the system becomes:

\begin{figure}[h!]
	\centering
	\includegraphics[width=0.23\textwidth, height =0.20\textwidth]{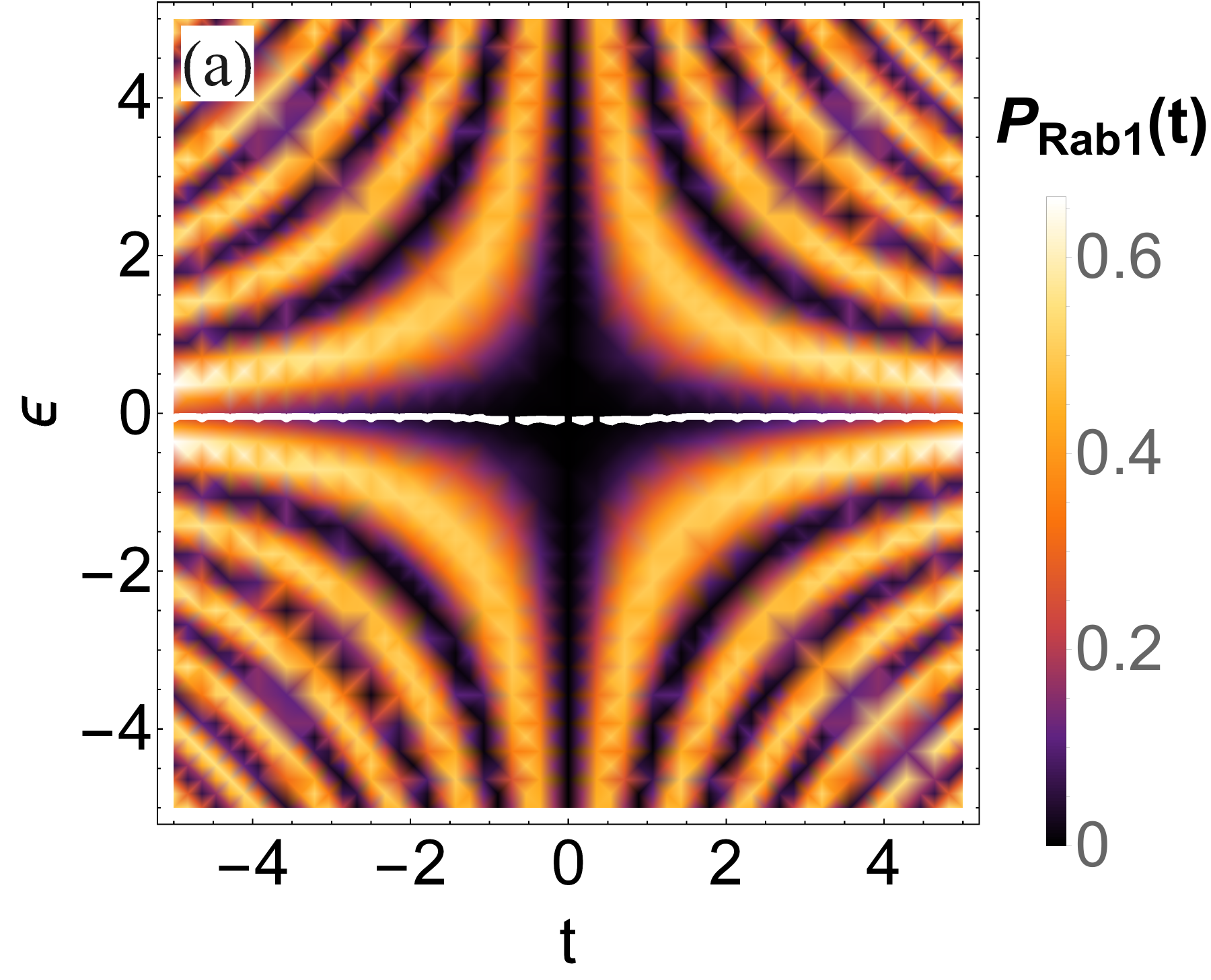}
	\includegraphics[width=0.20\textwidth, height =0.20\textwidth]{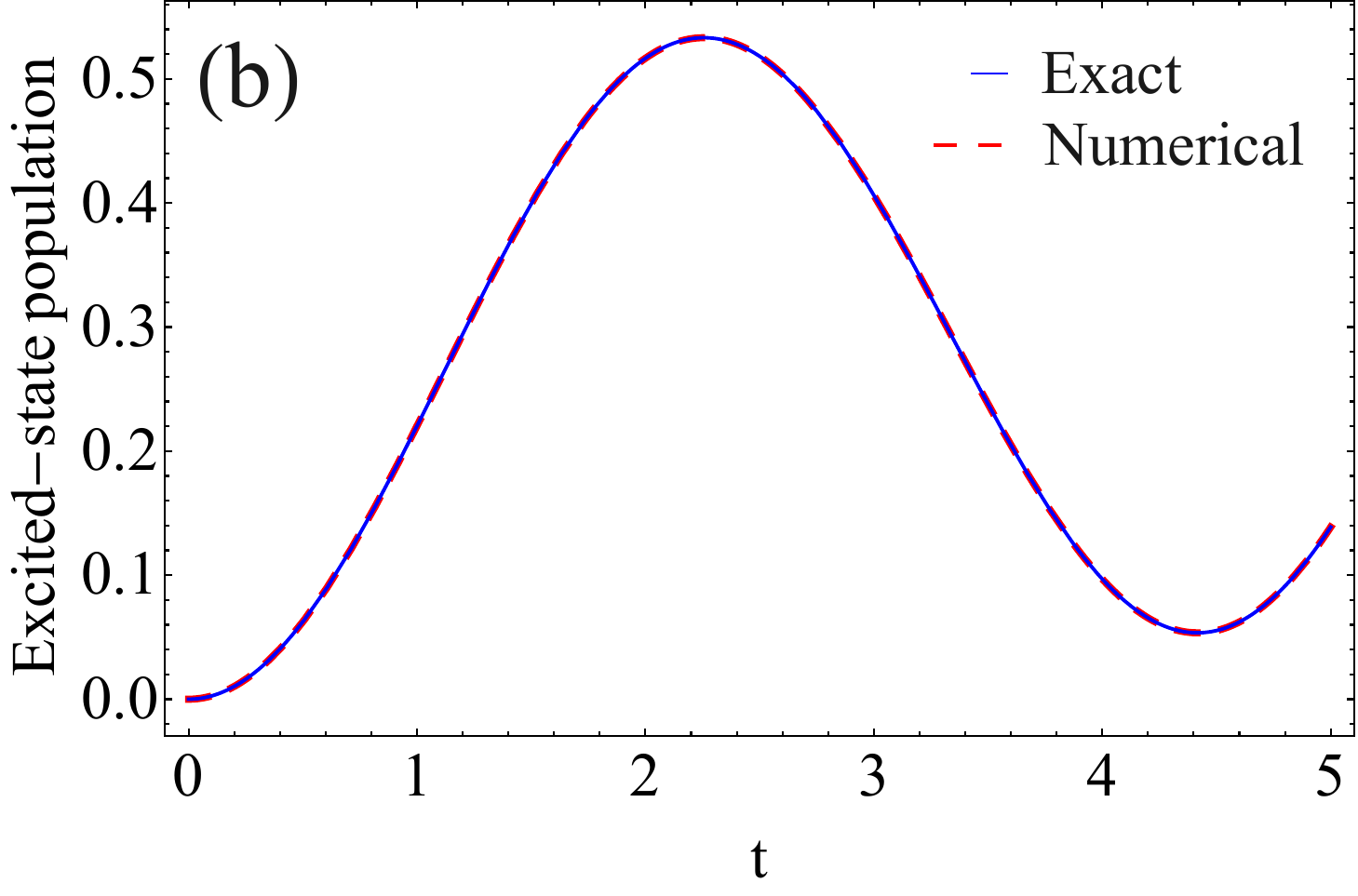}
	\caption{\small (Color online) a) Interferogram of the population versus the time and the shift, we remark the presence of many waves around the crossing region at $\epsilon = 0$ , we can assimilate the white line at $\epsilon = 0$ as a mirror that reflects any waves in the opposite direction. b) Similarity between the exact results and the numerical solutions against time of the excited-state population. We consider $\Delta = 0.2$}\label{fig7}
\end{figure}

\begin{equation}
	\frac{{{d^2}{\psi _{1,2}}\left( t \right)}}{{d{t^2}}} \mp i\epsilon \frac{{d{\psi _{1,2}}\left( t \right)}}{{dt}} + \frac{1}{4}{\left( {i\Delta  + \epsilon } \right)^2}{\psi _{1,2}}\left( t \right) = 0.
\end{equation}

The initial conditions of the Rabi problem ${\psi _1}\left( 0 \right) = 1$ and ${\psi _2}\left( 0 \right) = 0$, help solve this differential equation and determine the survival and transition probabilities. The survival probability is given by:
\begin{equation}
	{P_{Rab1} (t)} = \frac{{4{\epsilon ^2} + 3{{\left( {i\Delta  + \epsilon } \right)}^2}}}{{4\left( {{\epsilon ^2} + {{\left( {i\Delta  + \epsilon } \right)}^2}} \right)}}{\sin ^2}\left( {\frac{{\sqrt {{\epsilon ^2} + {{\left( {i\Delta  + \epsilon } \right)}^2}} }}{2}t} \right).
\end{equation}

In Fig.\ref{fig7}, we plot the variation of the population as a function of time. In frame (a), the interferogram of the population shows the presence of multiple waves around a mirror at $\epsilon = 0$, which reflects them in the opposite direction. These waves are generated by the small values of the coupling, leading us to conclude that the coupling is responsible for the interferometry observed in the system. Interferometry allows the extraction of significant information about the system.

\section*{Conclusion}

We have presented the first exponential Nikitin model under the influence of spontaneous emission. This emission is accounted for by the presence of an imaginary coupling combined with a shift. The inclusion of the imaginary coupling introduces nonlinearity into the system's evolution and contributes to its non-Hermitian nature. To study the dynamics of this model, we determined the transition and survival probabilities as functions of time, coupling, and shift.

When the population varies with time, as shown in Fig.\ref{fig2}, we observed that the imaginary coupling and the shift enhance the transmission of information for small values in the ground-state population. Conversely, these parameters have no significant influence on the excited-state population. When populations vary with the coupling, as illustrated in  Fig.\ref{fig3}, both ground and excited states exhibit exponential decay. Rapid decay occurs only when the phase takes a high value. Similarly, Fig.\ref{fig4} demonstrates exponential decay in populations, but some oscillations appear when the phase is large, indicating information transfer.

The non-Hermitian nature of our system results in complex eigenenergies, which can be separated into real and imaginary parts. The evolution of these eigenenergies indicates that energy gain is localized over a broad region, represented by the real part of the energy. Meanwhile, information loss is confined to a smaller region, represented by the imaginary part. Furthermore, the imaginary part contributes significantly to the transmission of the Rabi population, creating a "mirror" effect that reflects waves and generates oscillations.

For future work, it would be interesting to analyze spontaneous emission at different sweep velocities, utilizing both the Demkov-Kunike and the second Nikitin model.

\section*{Acknowledgments}

A. D. Kammogne thanked M. B. Kenmoe for interesting scientific comments and the University of Dschang for its warm hospitality during this project.

\section*{Declaration}
This paper has received no funding or financial support from anyone.

\end{document}